\documentclass[aps, showpacs,superscriptaddress,nofootinbib]{revtex4-1}
\usepackage{ulem,graphicx,epsfig,amsmath,amsfonts,amssymb,xcolor,slashed}
\usepackage{lipsum}
\usepackage{epsfig}
\usepackage{epstopdf}
\usepackage{hyperref}
\usepackage{float}
\usepackage{wrapfig}
\usepackage{tikz-feynman}
\usepackage{environ}

\newcommand{\Mbare}{\stackrel{\circ}{M}_{\rm CQM}}

\newcommand{\Slash}[1]{{\ooalign{\hfil/\hfil\crcr$#1$}}}

\renewcommand\sout{\bgroup \color[rgb]{1,0,0} \ULdepth=-.5ex \ULset}
\allowdisplaybreaks[1]
\begin{document}
\title{On the nature of the lowest-lying odd parity charmed baryon
   $\Lambda_c(2595)$ and $\Lambda_c(2625)$ resonances}
\date{\today}

\author{J.~Nieves}
\affiliation{Instituto~de~F\'{\i}sica~Corpuscular~(centro~mixto~CSIC-UV),
  Institutos~de~Investigaci\'on~de~Paterna, Aptdo.~22085,~46071,~Valencia,
  Spain}

\author{R.~Pavao}
\affiliation{Instituto~de~F\'{\i}sica~Corpuscular~(centro~mixto~CSIC-UV),
  Institutos~de~Investigaci\'on~de~Paterna, Aptdo.~22085,~46071,~Valencia,
  Spain}
\begin{abstract} 

We study the structure of the $\Lambda_c(2595)$ and $\Lambda_c(2625)$
resonances in the framework of an effective field theory consistent with 
heavy quark spin and chiral symmetries, that incorporates the interplay between $\Sigma_c^{(*)}\pi-ND^{(*)}$ baryon-meson degrees of freedom and bare P-wave $c\bar ud$ quark-model states.  We  show that these two resonances are not HQSS partners. The $J^P= 3/2^-$ $\Lambda_c(2625)$ should be viewed mostly as a dressed three quark state, whose origin is determined by a bare state, predicted to lie very close to the mass of the  resonance. The $J^P= 1/2^-$ $\Lambda_c(2595)$  seems to have, however,  a predominant molecular structure. This is because, it  is either the result of the chiral $\Sigma_c\pi$ interaction, which threshold is located much more closer than the mass of the bare three-quark state, or because the light degrees of freedom in its inner structure are coupled to the unnatural $0^-$ quantum-numbers. We show that both situations can occur depending on the renormalization procedure used. We find some additional states, but the  classification of the spectrum in terms of HQSS is difficult, despite having  used interactions that respect this symmetry. This is because  the bare quark-model state and the $\Sigma_c\pi$ threshold are located extraordinarily close to the $\Lambda_c(2625)$ and $\Lambda_c(2595)$, respectively, and hence they play totally different roles in each sector.

\end{abstract}
\maketitle

\section{Introduction}
\label{sec_introduction}

In the infinite quark mass limit ($m_Q\to \infty$), the dynamics of baryons containing a heavy quark should show a
  SU(2)--pattern, because of the symmetry that Quantum Chromodynamics (QCD)  acquires in that
  limit under arbitrary rotations of the spin of the heavy quark, $S_Q$. This
  is known as heavy quark spin symmetry (HQSS)~\cite{Isgur:1991wq,Wise:1992hn,Neubert:1993mb}. The
  total angular momentum $j_{ldof}$ of the light degrees of freedom ({\it ldof}) inside of the hadron is conserved, and heavy baryons with $J=j_{ldof}\pm1/2$ are expected to form a degenerate doublet.

  Constituent quark models (CQMs) find a nearly degenerate pair of
  $P-$wave $\Lambda_c^*$ excited states, with spin--parity $J^P=1/2^-$
  and $3/2^-$, and masses  similar to those of the isoscalar
  odd-parity $\Lambda_c(2595)$ and $\Lambda_c(2625)$
  resonances~\cite{Copley:1979wj,Migura:2006ep,Garcilazo:2007eh,Roberts:2007ni,Yoshida:2015tia}.  Two different
  excitation-modes are generally considered.  In the 
  $\lambda-$mode, excitations between the heavy quark and
  the {\it ldof} are accounted for, while in the $\rho-$mode,
  excitations in the inner structure of the {\it ldof} are instead considered. For singly-heavy baryons,
  the typical excitation energies of the $\lambda-$mode are smaller than those of the $\rho-$mode~\cite{Isgur:1991wr, Yoshida:2015tia}.  Within this picture, the
  $\Lambda^{\rm CQM}_c(2595)$ and $\Lambda^{\rm CQM}_c(2625)$ resonances would correspond
  to the members of the HQSS--doublet associated to
  $(\ell_\lambda=1,\ell_\rho=0)$, with total spin $S_q=0$ for the {\it ldof}. The total spins of
  these states are the result of coupling the orbital-angular
  momentum $\ell_\lambda$ of the {\it ldof}--with respect to the heavy
  quark-- with $S_Q$. Therefore,  both
  $\Lambda^{\rm CQM}_c(2595)$ and $\Lambda^{\rm CQM}_c(2625)$ states are connected by
  a simple rotation of the heavy-quark spin, and these resonances will be degenerate in
  the $m_Q\to \infty$ limit.

  Since the total angular momentum and parity of the {\it ldof} in the $S-$wave $\Sigma_c\pi$ and $\Sigma^*_c\pi$ pairs are
  $1^-$, as in the CQM  $\Lambda_c(2595)$ and $\Lambda_c(2625)$ resonances, the $\Lambda^{\rm CQM}_c(2595,2625) \to\pi\Sigma{(*)}_c  \to  \pi\pi\Lambda_c$ 
  decays respect HQSS, and hence   one could expect sizable widths for these
  resonances, unless these   transitions are  kinematically
  suppressed. This turns out to be  precisely the case~\cite{Tanabashi:2018oca}, and as it is shown  in Refs.~\cite{Nagahiro:2016nsx, Arifi:2017sac}, the use of the actual resonance masses lead to widths
  for the CQM $(\ell_\lambda=1,\ell_\rho=0)$ states ($j_{ldof}^\pi=1^-$)  predicted in
  \cite{Yoshida:2015tia} consistent with data.
  
  Within CQM schemes, it is nevertheless unclear why the role played by the $\Sigma_c^{(*)}\pi$ baryon-meson pairs in the generation of the $\Lambda_c(2595)$ and $\Lambda_c(2625)$ resonances can be safely ignored, especially in the $\Lambda_c(2595)$ case, since it is located very close to the $\Sigma_c\pi$ threshold (1 MeV below or four MeV above depending on the charged channel). This observation leads us naturally to consider molecular descriptions of these lowest-lying odd parity charmed baryon states, which should show up as poles in coupled-channel $T-$matrices, fulfilling exact unitary.  
  
  The first molecular studies~\cite{Lutz:2003jw, Tolos:2004yg} of the $\Lambda_c(2595)$ and $\Lambda_c(2625)$  were motivated by the appealing similitude of these resonances to the $\Lambda(1405) $ and $ \Lambda(1520) $ in the strange sector. In
particular the two isoscalar $S$-wave $\Lambda(1405)$ and
$\Lambda_c(2595)$ resonances have several features in common. The mass
of the former lies in between the $\Sigma\pi$ and $N \bar K $ 
thresholds, to which it couples strongly.  In turn, the
$\Lambda_c(2595)$ lies below the $ND$ and just slightly above the
$\Sigma_c\pi$ thresholds, and substituting the $c$ quark by a $s$
quark, one might expect the interaction  of $ND$ to play a role in the dynamics of the $\Lambda_c(2595)$ similar to that played by 
$N \bar{K}$ in the strange sector.  The
first works had some clear limitations. The $J^P= 1/2^-$ sector was studied in  \cite{Lutz:2003jw}, where 
the amplitudes obtained from  the scattering of Goldstone-bosons  off $1/2^+$ heavy-light baryons were unitarized. Despite
the interactions being fully consistent with chiral symmetry,  neither the $N D$, nor the $N D^*$
channels were  considered\footnote{A detailed treatment of the interactions between the ground-state $1/2^+$ and $3/2^+$ singly charmed and bottomed baryons 
and the pseudo-Nambu-Goldstone bosons, discussing also  the effects of the next-to-leading-order chiral potentials, can be found in \cite{Lu:2014ina}.}.  The work of Ref.~\cite{Tolos:2004yg} also studied the $\Lambda_c(2595)$ and there, the
interactions were obtained from chirally motivated Lagrangians upon
replacing the $s$ quark by the $c$ quark. Though in this way, the $N D$ channel was accounted for, the HQSS counterpart $N D^*$  was not considered.  
Subsequent works~\cite{Hofmann:2005sw, Mizutani:2006vq, Hofmann:2006qx, JimenezTejero:2009vq} introduced some
improvements, but they  failed to provide a scheme fully consistent with HQSS. In all cases, the $\Lambda_c(2595)$,  or  the $\Lambda_c(2625)$ when studied, could be dynamically generated after a convenient tuning of the low energy constants (LEC) needed to renormalize the ultraviolet (UV) divergences resulting from the baryon-meson loops. As mentioned before, none of these works were consistent with HQSS since none of them considered the $N D^*$ channel~\cite{GarciaRecio:2008dp}. Heavy
pseudoscalar and vector mesons should be treated on equal footing,
since they are degenerated in the heavy quark limit, and  are
connected by a spin-rotation of the heavy quark that leaves unaltered
the QCD Hamiltonian in that limit. 

The first molecular description of the
$\Lambda_c(2595)$ and $\Lambda_c(2625)$ resonances, using interactions fully consistent with HQSS, was proposed in Refs.~\cite{GarciaRecio:2008dp, Romanets:2012hm}. In these works a consistent ${\rm SU(6)}_{\rm lsf} \times {\rm SU(2)}_{\rm HQSS}$
extension of the chiral Weinberg-Tomozawa (WT) $\pi N$ Lagrangian --where ``lsf'' stands for
light-spin-flavor symmetry--, was derived.  Two
states with $J^P =1/2^-$ were dynamically generated in the region of 2595 MeV. The first one, identified with the
$\Lambda_c(2595)$ resonance, was narrow and it strongly coupled to 
$N D$ and especially to $N D^*$, with a small coupling to the open
$\Sigma_c\pi$ channel.  Its wave-function had a large $j_{ldof}^\pi=0^-$ component that  coupled to the spin ($S_Q=\frac12$) of the charm
  quark gives a total $J^P = \frac12^-$ for the $\Lambda_c(2595)$. Since the transition of the dominant $j_q^P=0^-$
term of the $\Lambda_c(2595)$ to the final $\Sigma_c\pi$ state is forbidden by HQSS, this mechanism will act in addition to
any possible kinematical suppression.

The second $J^P =1/2^-$ state found in ~\cite{GarciaRecio:2008dp, Romanets:2012hm} was quite broad since it had a
sizable coupling to the $\Sigma_c\pi$ channel, and reproduced, in the charm-sector, the chiral two-pole structure of the $\Lambda(1405)$~\cite{Oller:2000fj,GarciaRecio:2002td,Hyodo:2002pk,Jido:2003cb,GarciaRecio:2003ks,Hyodo:2011ur,
  Kamiya:2016jqc}. On the other hand, a
$J^P=3/2^-$ state is generated mainly by the $(N D^*-\Sigma_c^*\pi)$
coupled-channel dynamics. It would be the charm counterpart of the
$\Lambda(1520)$, and it was argued that could be identified with the $\Lambda_c(2625)$ resonance.

Several $\Lambda_c^*$ poles were also obtained in the approach followed in
Ref.~\cite{Liang:2014kra}. There, the interaction of $N D$ and $N D^*$ states, together with their coupled
channels are considered by using an extension  of the SU(3) local hidden gauge formalism  from the light meson sector~\cite{Bando:1984ej,Bando:1987br,
  Meissner:1987ge} to four flavours . The scheme also respects LO HQSS constraints~\cite{Xiao:2013yca} and, as in Refs.~\cite{GarciaRecio:2008dp,
  Romanets:2012hm}, a two-pole structure for the $\Lambda_c(2595)$ was also found, with the $N D^*$ channel playing a crucial role in
its dynamics. This is a notable
difference to the situation in the strange sector, where the analog
$N \bar K^*$ channel is not even considered in most of the studies of
the $\Lambda(1405)$, because of the large $\bar K^*-\bar K$ mass
splitting. We will refer to this model as ELHG for the rest of this work.

However neither the ${\rm SU(6)}_{\rm lsf} \times {\rm SU(2)}_{\rm HQSS}$ model, nor the ELHG consider 
the interplay between $\Sigma_c^{(*)}\pi-ND^{(*)}$ baryon-meson degrees of freedom and bare P-wave $c\bar ud$ quark-model states. This is unjustified, in the same way,  it was also unjustified the neglect of baryon-meson effects  in the CQM approaches. 

The CQM approach of Ref.~\cite{Yoshida:2015tia} finds isoscalar $J^P=1/2^-$ and $3/2^-$states  at 2628 and 2630 MeV, respectively. Given the proximity of these bare three-quarks states to the $\Lambda_c(2595)$ and $\Lambda_c(2625)$, it is reasonable to expect
a significant influence of the CQM degrees of freedom on the dynamics of the physical states. This seems to be specially truth for the $\Lambda_c(2625)$, for which the CQM prediction almost matches its mass.  CQM degrees of freedom can be taken into account in hadron scattering schemes by considering an additional  
energy dependent interaction~\cite{Baru:2010ww, Cincioglu:2016fkm}, driven by a pole in the baryon-meson tree-level amplitudes located
at the bare mass, $\Mbare$,  of the CQM state. At energies far enough from $\Mbare$, the contribution of the CQM degrees of freedom can be possibly  accounted for an appropriate LEC (induced by the UV regulator of the loops)  in the unitarized baryon-meson amplitude. However, such contribution becomes more important for energies approaching $\Mbare$, and its energy dependence might then not be safely ignored. 

In this work we will study of the structure of the $\Lambda_c(2595)$ and $\Lambda_c(2625)$
resonances, in the framework of an effective theory consistent with 
heavy quark spin and chiral symmetries, incorporating for the very first time,  
the interplay between $\Sigma_c^{(*)}\pi-ND^{(*)}$ baryon-meson degrees of freedom and bare P-wave $c\bar ud$ quark-model states. For simplicity, we will use the ${\rm SU(6)}_{\rm lsf}\times$HQSS baryon-meson amplitudes, though the most important conclusions extracted here do not depend on the particular hadron scattering model employed. 

The work is organized as follows. After this Introduction, the used formalism is briefly revised in Sect.~\ref{sec:form}, that is split in several subsections dealing with the ${\rm SU(6)}_{\rm lsf}\times$HQSS hadron amplitudes, their renormalization and structure in the complex plane,  with the inclusion of the CQM degrees  of freedom and finally with the evaluation of the $\Lambda^*_c(1/2^-,3/2⁻) \to \Lambda_c(1/2^+)\pi\pi$ three-body decays. The results of this research are presented and discussed in Sect.~\ref{sec:res}, first neglecting CQM effects (Subsec.~\ref{sec:Su6results}) and next coupling CQM and baryon-meson degrees of freedom (Subsec.~\ref{sec:CQM-hadron}). Finally, the main conclusions of this work are summarized in Sect.~\ref{sec:concl}.

\section{Formalism}
\label{sec:form}
\subsection{${\rm SU(6)}_{\rm lsf}\times$HQSS amplitudes}

The building-blocks considered in  ~\cite{GarciaRecio:2008dp, Romanets:2012hm, Garcia-Recio:2013gaa} in the $C=1$ sector 
are the pseudoscalar ($D_s, D,K, \pi,\eta, {\bar K}, {\bar D}, {\bar D}_s$) and vector ($D_s^*,
D^*,K^*, \rho,\omega, {\bar K}^{*}, {\bar D}^*, {\bar D}_s^*, \phi$)
mesons, the spin--$1/2$ octet and the spin--$3/2$ decuplet of
low-lying light baryons, in addition to the spin-1/2 ($\Lambda_c$,
$\Sigma_c$, $\Xi_c$, $\Xi'_c$, $\Omega_c$), and spin-3/2
($\Sigma^*_c$, $\Xi^*_c$, $\Omega_c^*$) charmed
baryons. All baryon-meson pairs with
$(C=1, S=0, I=0)$ quantum numbers span the coupled-channel space for
a given total angular momentum $J$ and odd parity.  The $S$-wave tree level
amplitudes between two channels are given by the ${\rm SU(6)}_{\rm lsf}$ $\times$ HQSS
WT  kernel
\begin{equation}
\label{eq:WT}
V_{ij}^J(s) = D_{ij}^J \frac{2 \sqrt{s}-M_i-M_j}{4 f_i f_j} \sqrt{\frac{E_i+M_i}{2 M_i}} \sqrt{\frac{E_j+M_j}{2M_j}},
\end{equation}
with $s$ the baryon-meson Mandelstam variable,  $M_i$ and $m_i$, the masses of the baryon and meson in the $i$
channel, respectively, and $E_i =(s-m_i^2+M_i^2)/2 \sqrt{s}$, the center-of-mass
energy of the baryon in the same channel. The hadron masses and meson decay constants, $f_i$,  have been taken from
Ref.~\cite{Romanets:2012hm}. The $D_{ij}^J$ matrices are determined by
the underlying ${\rm SU(6)}_{\rm lsf} \times$ HQSS group
structure of the interaction. Tables for all of them can be found in the Appendix B of Ref.~\cite{Romanets:2012hm}. Here, we truncate the 
coupled-channels space  to that generated by the $\Sigma_c\pi$, $ND$ and $ND^*$ and $\Sigma_c^*\pi$ and $ND^*$ in the $J^P=1/2^-$ and $3/2^-$ sectors, 
respectively.  Other higher channels like $\Lambda_c \eta$, $\Lambda_c \omega $, $ \Xi^{(\prime, *)}_c K^{(*)} $, $ \Lambda D_s^{(*)}$, $\Sigma_c^{(*)} \rho, \ldots$ are little relevant for the  dynamics of the $\Lambda_c(2595)$ and $\Lambda_c(2625)$ resonances~\cite{GarciaRecio:2008dp, Romanets:2012hm}, and have not been considered in the analysis carried out in this work. 

The matrices $D^J$ are given in ~\cite{Romanets:2012hm} in a basis of $S-$wave baryon-meson states. They become, however,  diagonal when states with well defined {\it ldof} total angular momentum, $j_{ldof}$, are used. For the latter states, HQSS constrains are straightforward  because of the symmetry that QCD acquires, in the infinite quark mass limit,  under arbitrary rotations of the spin of the heavy quark ~\cite{Isgur:1991wq,Wise:1992hn,Neubert:1993mb}.   In both bases, the total angular momentum of the baryon-meson pair is defined, and both sets of states are related by a Racah rotation~\cite{Xiao:2013yca, Nieves:2019kdh}. The eigenvalues of $D^J$ in the $(\Sigma_c^{(*)}\pi, ND^{(*)})$ truncated space  are $\lambda_0=-12$, $\lambda_1^{\rm atr}=-2-\sqrt{6}$ and  $\lambda_1^{\rm rep}=-2+\sqrt{6}$, and  $\lambda_1^{\rm atr}$ and $\lambda_1^{\rm rep}$, for $J^P=1/2^-$ and $J^P=3/2^-$, respectively~\cite{Nieves:2019kdh}. Actually, the ${\rm SU(6)}_{\rm lsf} \times {\rm SU(2)}_{\rm HQSS}$
extension of the WT $\pi N$ interaction proposed in ~\cite{GarciaRecio:2008dp, Romanets:2012hm, Garcia-Recio:2013gaa} leads to a large attraction $(\lambda_0)$ in the  
subspace where the total angular-momentum--parity quantum numbers of the {\it ldof} are $j^\pi_{ldof}=0^-$. This latter configuration does not occur for $J=3/2$, when only $S-$wave interactions are considered, and the  {\it ldof} are necessarily coupled to $j^\pi_{ldof}=1^-$. In the $j^\pi_{ldof}=1^--$subspace, there exist both  attractive ($\lambda_1^{\rm atr}$) and repulsive ($\lambda_1^{\rm rep}$) components, and HQSS relates the $D^{J=1/2}$ and $D^{J=3/2}$ matrices. 
\subsection{Renormalization of the Bethe--Salpeter equation}
\label{sec:renor}

We use the matrix $V_{ij}^J$ as kernel to solve the Bethe-Salpeter
equation (BSE), leading to a $T$-matrix 
\begin{equation}
\label{eq:LS}
T^J(s)=\frac{1}{1-V^J(s) G^J(s)} V^J(s),
\end{equation}
satisfying exact unitarity in coupled channels. The diagonal matrix $G^J(s)$ is constructed out of the baryon-meson loop functions
\begin{equation}
\label{eq:normloop}
G_i(s)=i 2M_i  \int \frac{d^4 q}{(2 \pi)^4} \frac{1}{q^2-m_i^2+i\epsilon} \frac{1}{(P-q)^2-M_i^2+i\epsilon},
\end{equation}
with $P$ the total momentum of the system such that $P^2=s$. We omit
the index $J$ from here on for simplicity. The bare loop function is
logarithmically UV divergent and needs to be
renormalized. This can be done by one-subtraction
\begin{equation}
G_i(s)=\overline{G}_i(s)+G_i(s_{i+}) ,
\label{eq:div}
\end{equation}
where the finite part of the loop function, $\overline{G}_i(s)$, reads 
\begin{equation}
\overline{G}_i(s) = \frac{2 M_i}{(4 \pi)^2} \left(\left[ \frac{M_i^2-m_i^2}{s}-\frac{M_i-m_i}{M_i+m_i}\right] \log \frac{M_i}{m_i}+L_i(s) \right),\label{eq:defG}
\end{equation}
with $s_{i\pm}=(m_i\pm M_i)^2$, and the multi-valued function $L(s)$ given in Eq.~(A10) of Ref.~\cite{Nieves:2001wt}.

The divergent contribution of the loop function, $G_i(s_{i+})$ in
Eq.~(\ref{eq:div}) needs to be renormalized.  We will
examine here two different renormalization schemes. On the one hand, we will  perform
one subtraction at certain scale $\sqrt{s}=\mu$, such that $G_i(\sqrt{s}=\mu) = 0$.  In this way,
\begin{equation}
G_i^\mu(s_{i+})  = - \overline{G}_i(\mu^2).
\end{equation}
In addition,  we consider the prescription employed in
Refs~\cite{GarciaRecio:2008dp, Romanets:2012hm}, where a common scale $\mu$ is chosen to be  independent
of the total angular momentum $J$~\cite{Hofmann:2005sw, Hofmann:2006qx}, and equal to
\begin{equation}
  \mu = \sqrt{\alpha \left(m_\pi^2+M_{\Sigma_c}^2 \right)}\label{eq:defmualpha} ,
\end{equation}
in the sectors of the $\Lambda_c(2595)$ and $\Lambda_c(2625)$ resonances. In the equation above, $\alpha$ is a parameter that can be adjusted to
data~\cite{GarciaRecio:2008dp}. In what follows, we will refer to this
scheme as SC$\mu$ (subtraction at common scale).
 
In the second  renormalization scheme, we make finite the UV divergent part of
the loop function using a cutoff regulator $\Lambda$ in momentum space,
which leads to~\cite{GarciaRecio:2010ki}
\begin{eqnarray}
G_i^\Lambda(s_{i+}) &=& \frac{1}{4\pi^2} \frac{M_i}{m_i+M_i} \left
(m_i\ln\frac{m_i}{\Lambda + \sqrt{\Lambda^2+m_i^2}}+  M_i\ln\frac{M_i}{\Lambda + \sqrt{\Lambda^2+M_i^2}} \right)\label{eq:uvcut} ,
\end{eqnarray}
Note that,  there are no cutoff effects in the  finite
$\overline{G}_i(s)-$loop function,  as it would happen if the two-body
propagator of Eq.~\eqref{eq:normloop} would have been directly calculated using
the UV cutoff $\Lambda$.

If a common UV cutoff is used for all channels, both renormalization schemes are independent and will lead to
different results. However, if one allows the freedom of using
channel-dependent cutoffs, the subtraction a common scale scheme, SC$\mu$,  is recovered by
choosing in each channel, $\Lambda_i$ such that
\begin{equation}
G^{\Lambda_i}_i(s_{i+})= -\overline{G}_i(\mu^2) .
\label{eq:subtraction}
\end{equation}
\subsection{Interplay between bare CQM and baryon-meson degrees of freedom}
\label{sec:QMth}
 Within the quark model approach of Ref.~\cite{Yoshida:2015tia}, odd-parity $\Lambda^*$ states are obtained at 2628 and 2630 MeV for $J=1/2$ and 3/2, respectively. The {\it ldof} are coupled to angular momentum--parity quantum numbers $j^\pi_{ldof}=1^-$ ($\lambda-$mode) in both cases, which explains their approximate degeneracy. Higher excited states appear at 2.9 GeV ($\rho-$mode), far from the $\Lambda_c(2595)$ and $\Lambda_c(2625)$ narrow resonances, and will not be considered in the present analysis. However. the low-lying $\lambda-$mode states, given their proximity to the $\Lambda_c(2595)$ and $\Lambda_c(2625)$, might significantly influence the dynamics of the physical states. This seems to be specially truth for the $\Lambda_c(2625)$, since the prediction of Ref.~\cite{Yoshida:2015tia} for its mass is only 2 MeV higher than the experimental one [$(2628.11 \pm 0.19)$ MeV~\cite{Tanabashi:2018oca}].
 
 Bare CQM-states effects on the baryon-meson dynamics can be effectively considered by means of an energy dependent interaction~\cite{Cincioglu:2016fkm, Albaladejo:2016ztm, Albaladejo:2018mhb}
\begin{equation}
 \label{eq:exchange}
 V_\text{ex}^J(s)=2\Mbare\frac{[V^J_{\rm CQM}]^\dagger \cdot [V^J_{\rm CQM}]}{s-(\Mbare)^2}\,,\qquad  V^{J=1/2}_{\rm CQM} = \left (\begin{array}{ccc}d_1 & \sqrt{3}c_1/2
 & -c_1/2 \cr
 \Sigma_c\pi & ND & ND^*
 \end{array}
 \right) \,,\qquad  V^{J=3/2}_{\rm CQM} = \left (\begin{array}{ccc}-d_1 & c_1 \cr
 \Sigma_c^*\pi & ND^*
 \end{array}
 \right) 
\end{equation}
where $d_1$ and $c_1$ are  undetermined dimensionless parameters  that control the strength of the baryon-meson-CQM-state vertex. The above interaction accounts for the contribution to baryon-meson scattering of the exchange of an intermediate odd-parity CQM $\lambda-$mode state. It does not  obviously affect the $j^\pi_{ldof}=0^--$subspace of the $J^P=1/2^-$ sector, and it is consistent with HQSS in the $j^\pi_{ldof}=1^-$subspace of  the $J^P=1/2^-$ and $J^P=3/2^-$ sectors, which are related by a spin rotation of the heavy quark. 

Note that $ V_\text{ex}^J(s)$ introduces a pole in the baryon-meson tree-level amplitudes located at the bare mass value, $\sqrt{s}=\Mbare$. It should be interpreted as the mass of the CQM state in the limit of vanishing coupling to the  baryon--meson-pairs ($d_1,c_1\to 0$), and therefore it is not an observable.   The interaction with the baryon-meson cloud \textit{dresses} the CQM state through loops, renormalizing its mass, and the dressed  state might also acquire a finite width, when it is located above threshold.  A priori, $\Mbare$ is a free parameter of the present approach, and moreover it depends on the renormalization scheme~\cite{Cincioglu:2016fkm}. This is because, in the effective theory, the UV regulator is finite, and the difference between the bare and the physical resonance masses is a finite renormalization that depends on the adopted scheme.  The value of the bare mass, which is  thus a free parameter, can either be indirectly fitted to
experimental observations, or obtained from schemes that ignore the  coupling to  baryon-meson pairs, such as some CQMs. In this latter case, the issue certainly
  would be to set the UV regulator to match the quark model and the  baryon-meson scattering approaches~\cite{Cincioglu:2016fkm}. For simplicity, and consistently with HQSS, we take a common bare mass  for both $J=1/2$ and $J=3/2$, which is fixed to the average of masses reported in the quark model of Ref.~\cite{Yoshida:2015tia} ($\Mbare=2629$ MeV). We will explore different values of the renormalization scheme-dependent bare couplings $d_1$ and $c_1$ to elucidate the robustness of our results.

At energies far enough from $\Mbare$, the contribution of $V_{\rm ex}$ can be regarded as a small contact interaction  
that can be accounted for by means of a LEC. However, the exchange contribution becomes more important for energies approaching $\Mbare$, and it may not be safe to ignore its energy dependence. One might expect such situation in the  $J=3/2$ sector, where  $V_{\rm ex}$ should  provide a sizable attraction (repulsion) for energies slightly below (above) $\Mbare$, relevant in the dynamics of the $\Lambda_c(2625)$. We expect a less relevant role in the case of the $\Lambda_c(2595)$, since this resonance is located furthest from $\Mbare$.  

\subsection{Riemann sheets, poles and residues}
Masses and widths of the dynamically generated resonances in each
$J-$ sector are determined from the positions of the poles, $\sqrt{s_R}$, in
the second Riemann sheet (SRS) of the corresponding baryon-meson scattering
amplitudes, namely $\sqrt{s_R}= M_R-i\, \Gamma_R/2$. In some cases, we
also find real poles in the first Riemann sheet (FRS) which correspond to bound states. The different Riemann sheets  are defined in terms of the multivalued function $L(s)$, introduced in Eq.~\eqref{eq:defG}, that is evaluated here as explained in Eq.~(A13) of Ref.~\cite{Nieves:2001wt}, and thus are labeled by $\xi_1\xi_2\xi_3$ [$\xi_1\xi_2$], with $\xi_i=0,1$, in the $J=1/2\, [3/2]$ sector. The SRS in the relevant
fourth quadrant is obtained from the first quadrant FRS by continuity across each of the four unitarity cuts. 

The coupling constants of each resonance to the various baryon-meson
states are obtained from the residues at the pole 
by matching the amplitudes to the
expression
\begin{equation}
T^J_{ij}(s)=\frac{g_i  g_j }{\sqrt{s}-\sqrt{s_R}} \ , \label{eq:pole}
\end{equation}
for energy values $s$ close to the pole, where the dimensionless couplings, $g_i$,  
turn out to be in general complex.

\subsection{The $\Lambda^*_c(1/2^-,3/2⁻) \to \Lambda_c(1/2^+)\pi\pi$ three-body decays through the $\pi\Sigma^{(*)}_c$ intermediate state}
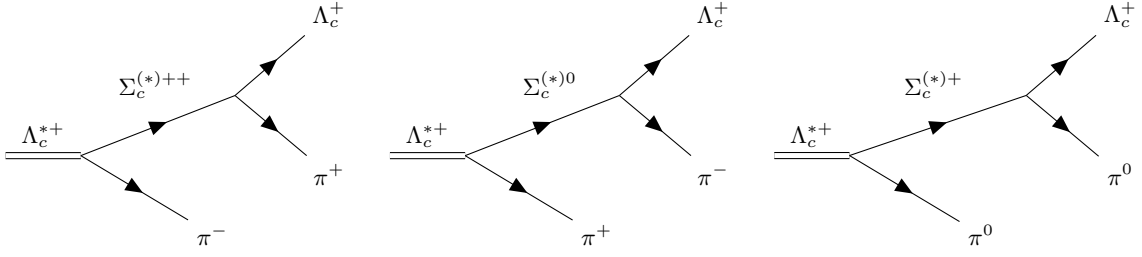
\begin{figure}
\scalebox{1.}{
\begin{tikzpicture}
\begin{feynman}
\vertex (v1);
\vertex[left= 1cm of v1](a1);
\vertex[right=1.75cm of v1](b1);
\vertex[right = .3cm of b1](b3);
\vertex[below =  0.8 cm of b1] (a2){$\pi^-$};

\vertex[above =  .8 cm of b3] (v2);
\vertex[right = 1.25cm of v2](b2);
\vertex[above = 0.8cm of b2](a4){$\Lambda_c^+$};
\vertex[below = 0.8cm of b2](a3){$\pi^+$};

\vertex[above = 2 cm of a2] (sl) ;
\vertex[left = .15 cm of sl] (sl2)  {$\Sigma_c^{(*)++}$};

\diagram* {
(a1)-- [ double distance=1.9pt, edge label = $\Lambda_c^{*+}$](v1)-- [fermion](v2)--[fermion](a4),
(v1)--[fermion](a2),
(v2)--[fermion](a3),
};
\end{feynman}
\end{tikzpicture}}
~
\scalebox{1.}{
\begin{tikzpicture}
\begin{feynman}
\vertex (v1);
\vertex[left= 1cm of v1](a1);
\vertex[right=1.75cm of v1](b1);
\vertex[right = .3cm of b1](b3);
\vertex[below =  0.8 cm of b1] (a2){$\pi^+$};

\vertex[above =  .8 cm of b3] (v2);
\vertex[right = 1.25cm of v2](b2);
\vertex[above = 0.8cm of b2](a4){$\Lambda_c^+$};
\vertex[below = 0.8cm of b2](a3){$\pi^-$};

\vertex[above = 2 cm of a2] (sl) ;
\vertex[left = .15 cm of sl] (sl2)   {$\Sigma_c^{(*) 0}$};

\diagram* {
(a1)-- [ double distance=1.9pt, edge label = $\Lambda_c^{*+}$](v1)-- [fermion](v2)--[fermion](a4),
(v1)--[fermion](a2),
(v2)--[fermion](a3),
};
\end{feynman}
\end{tikzpicture}}
~
\scalebox{1.}{
\begin{tikzpicture}
\begin{feynman}
\vertex (v1);
\vertex[left=1cm of v1](a1);
\vertex[right=1.75cm of v1](b1);
\vertex[right = .6cm of b1](b3);
\vertex[below =  0.8 cm of b1] (a2){$\pi^0$};

\vertex[above =  .8 cm of b3] (v2);
\vertex[right = 1.25cm of v2](b2);
\vertex[above = 0.8cm of b2](a4){$\Lambda_c^+$};
\vertex[below = 0.8cm of b2](a3){$\pi^0$};

\vertex[above = 2 cm of a2] (sl) ;
\vertex[left = .12 cm of sl] (sl2)   { $\Sigma_c^{(*)+}$};

\diagram* {
(a1)-- [ double distance=1.9pt, edge label = $\Lambda_c^{*+}$](v1)-- [fermion](v2)--[fermion](a4),
(v1)--[fermion](a2),
(v2)--[fermion](a3),
};
\end{feynman}
\end{tikzpicture}}
\caption{Diagrams for $\Lambda_c(2595)$ or $\Lambda_c(2625)$ decay into $\Lambda_c^+$ and two pions, mediated by the $\Sigma_c$ or $\Sigma^*_c$ resonances, respectively.}
\label{fig:decay}
\end{figure}
Isospin conservation forbids  single pion transitions between the $\Lambda_c^*$ and $\Lambda_c$, and hence the $\Lambda^*_c(1/2^-,3/2⁻)$ decay into $\Lambda_c$ and two pions. The decays proceed via an intermediate $I=1$ baryon-state down to a $\Lambda_c\pi$ pair.  The relatively small masses of the $\Lambda_c^{*}$'s above $\Lambda_c$ kinematically restrict the outgoing pion energies, making feasible a chiral derivative expansion~\cite{Cho:1994vg}.
There are two different final states, $\Lambda_c^+\pi^-\pi^+$ and $\Lambda_c^+\pi^0\pi^0$, and we will consider here only the resonant term driven by the excitation of the $\Sigma_c^{(*)}$ and its subsequent decay into  $\Lambda_c\pi$, as shown in Fig.~\ref{fig:decay}. This is by large the dominant contribution to the $\Lambda_c(2595)$ width, while it becomes significantly smaller for the $\Lambda_c(2625)$ one, since the virtual $\Sigma_c^*$ intermediate state is very much off-shell~\cite{Cho:1994vg,Pirjol:1997nh}. Indeed the ARGUS Collaboration reported a ratio~\cite{Albrecht:1993pt} $R=\Gamma[\Lambda_c(2625)\to \Lambda_c^+\pi^+\pi^-\,(\rm non-resonant)]/\Gamma[\Lambda_c(2625)\to \Lambda_c^+\pi^+\pi^-\,(\rm total)]= 0.54 \pm 0.14$.

The $\Lambda_c(2595)$ or $\Lambda_c(2625)$ decay width into the charged-pions mode\footnote{The $\Lambda(2595)$ and $\Lambda(2625)$  amplitudes read 
\begin{eqnarray}
 T_{\pi^-\pi^+}^{J=1/2} &=& - \frac{g_D g_{R\Sigma_c\pi}}{3f_\pi}\frac{2M_{\Sigma_c}}{(M_R-M_{\Sigma_c})} \bar u_{\Lambda_c}\left\{ (vp_{\pi^-})G_{\Sigma_c}(s_{12})(\Slash{p}_{\pi^+}+vp_{\pi^+})\gamma_5\frac{1+\Slash{v}}{2}+ (vp_{\pi^+})G_{\Sigma_c}(s_{13})(\Slash{p}_{\pi^-}+vp_{\pi^-})\gamma_5\frac{1+\Slash{v}}{2}\right\}u_{R} \\
 T_{\pi^-\pi^+}^{J=3/2} & = & \frac{g_D g_{R\Sigma_c\pi}}{3f_\pi}\frac{2M_{\Sigma_c}}{(M_R-M_{\Sigma_c})} \left[(vp_{\pi^-})G_{\Sigma_c}(s_{12})
 p_{\pi^+}^\mu + (vp_{\pi^+})G_{\Sigma_c}(s_{13})p_{\pi^-}^\mu \right] \bar u_{\Lambda_c} P_{\mu\nu}(v) u_R^\nu
\end{eqnarray}
where $v$ is the common velocity of all involved charmed hadrons, which remains unaltered in the heavy quark limit. It satisfies $v^2=1$ and $\Slash{v}u^{(\nu)}_{R}= u^{(\nu)}_R$ and $\Slash{v}u_{\Lambda_c}= u_{\Lambda_c}$, with $u$ and $u^\nu$ mass-dimensions Dirac and Rarita-Scwinger spinors, respectively. The spin 3/2 projector is 
\begin{equation}
 P^{\mu\nu}(v) = \left(-g^{\mu\nu}+v^\mu v^\nu + \frac13 (\gamma^\mu+v^\mu)(\gamma^\nu-v^\nu)\right) \frac{1+\Slash{v}}{2}
\end{equation}
with metric $(+,-,-,-)$, and the sum over fermion polarizations is given by $u\bar u = 2M(1+\Slash{v})/2$ and $u^\mu\bar u^\nu = 2M P^{\mu\nu}$.}, at lowest order in chiral perturbation theory and in the heavy mass limit,  is given in the resonance rest frame (LAB) by~\cite{Cho:1994vg}
\begin{eqnarray}
\frac{d\Gamma_{\pi^-\pi^+}}{ds_{12}ds_{23}}&=&\Gamma_0 \left\{ E_3^2 \vec{p}_2^{\,2}\left|G_{\Sigma_c^{(*)}}(s_{12})\right|^2+ E_2^2 \vec{p}_3^{\,2}\left|G_{\Sigma_c^{(*)}}(s_{13})\right|^2 + 2E_2E_3{\rm Re}\left[G^*_{\Sigma_c^{(*)}}(s_{12})G_{\Sigma_c^{(*)}}(s_{13})\right]\vec{p}_2\cdot\vec{p}_3\right\} \label{eq:chwdth}\\
\Gamma_0 &=& \frac{g_D^2 g^2_{R\Sigma_c^{(*)}\pi}}{144 f_\pi^2\pi^3}\frac{M_{\Lambda_c} M^2_{\Sigma_c^{(*)}} }{M_R^2 (M_R-M_{\Sigma_c^{(*)}})^2} \, \qquad G_{\Sigma_c^{(*)}}(s) = \frac{1}{s-M^2_{\Sigma_c^{(*)}}+ i\,M_{\Sigma_c^{(*)}}\Gamma_{\Sigma_c^{(*)}}(s)} \nonumber \\
\Gamma_{\Sigma_c^{(*)}}(s) &=& \frac{g_D^2}{6\pi f_\pi^2}\frac{M_{\Lambda_c}}{M_{\Sigma_c^{(*)}}}|\vec{p}_\pi^{\,\Lambda_c\pi}|^3\, ,\qquad |\vec{p}_\pi^{\,\Lambda_c\pi}|= \frac{\lambda^{\frac12}(s,M_{\Lambda_c}^2,m_\pi^2)}{2\sqrt{s}},\qquad s_{13}= M_R^2+2m_\pi^2+M_{\Lambda_c}^2-s_{12}-s_{23} \nonumber \\
E_3 &=&\sqrt{m_\pi^2+\vec{p}_3^{\,2}} = \frac{M_R^2+m_\pi^2-s_{12}}{2M_R}\, , ~ E_2 =\sqrt{m_\pi^2+\vec{p}_2^{\,2}} = \frac{M_R^2+m_\pi^2-s_{13}}{2M_R}\,, ~ \vec{p}_2\cdot\vec{p}_3 = E_2E_3+m_\pi^2-s_{23}/2\nonumber
\end{eqnarray}
with $M_R$ the resonance mass, $\lambda(x,y,z) = x^2 +y^2+z^2-2xy-2xz-2yz$. In addition,  $s_{12}$ (invariant mass square of $\Lambda_c\pi^+$) varies between $(M_{\Lambda_c}+m_\pi)^2$ and $(M_R-m_\pi)^2$, while the limits of $s_{23}$ (invariant mass square of the $\pi^+\pi^-$ pair) are
\begin{eqnarray}
s_{23}^{\rm max,\,min} = \left(E^*_{\pi^+}+E^*_{\pi^-}\right)^2-\left(p^*_{\pi^+}\mp p^*_{\pi^-}\right)^2 \, , \quad  
E^*_{\pi^+} = \frac{s_{12}+m_\pi^2-M_{\Lambda_c}^2}{2\sqrt{s_{12}}}\,, \quad E^*_{\pi^-} = \frac{M_R^2-m_\pi^2-s_{12}}{2\sqrt{s_{12}}}
\end{eqnarray}
with $p^{*2}_{\pi^\pm}= E^{*2}_{\pi^\pm}-m_\pi^2$. The expression of Eq.~\eqref{eq:chwdth} corresponds to the square of the sum of amplitudes associated to the first two diagrams of 
Fig.~\ref{fig:decay}. The processes occur so close to threshold, specially the $\Lambda_c(2595)$ decay,  that the available phase space might depend significantly on the small isospin-violating mass differences between members of the pion and $\Sigma_c^{(*)}-$multiplets. We have used $m_\pi=m_{\pi^\pm}$, $M_{\Sigma_c^{(*)}}= (M_{\Sigma_c^{++(*)}}+M_{\Sigma_c^{0(*)}} )/2$, $M_{\Lambda_c(2595)}= 2592.25$ MeV and $M_{\Lambda_c(2625)}= 2628.11$ MeV. The errors on the masses of the $\Lambda_c^*$ resonances quoted in the RPP~\cite{Tanabashi:2018oca} are 0.28 MeV and 0.19 MeV, respectively, and turn out to be  relevant only for the $\Lambda_c(2595)$ width, but even in that case, it induces variations of the order of 1\%. In addition $g_D/f_\pi=0.0074$ MeV$^{-1}$, which leads to $\Gamma[\Sigma_c\to \Lambda_c \pi]= 1.9$ MeV and 
$\Gamma[\Sigma_c^*\to \Lambda_c \pi]= 14.4$ MeV, and we take the dimensionless coupling $g_{R\Sigma_c^{(*)}\pi}$ from the residue at the resonance-pole (Eq.~\eqref{eq:pole}) of the $\Sigma_c^{(*)}\pi$ channel ($S-$wave) that we choose to be real, by an appropriate redefinition of the overall phases of the meson and baryon fields. 

The rates for the neutral-pions channel can be obtained by adding a symmetry factor 1/2 to avoid double counting the two identical bosons in the final state and using 
$m_\pi=m_{\pi^0}$, $M_{\Sigma_c^{(*)}}= M_{\Sigma_c^{+(*)}}$.

Adding the contribution of neutral and charged pion modes, we find that the $\Sigma_c^{(*)}-$resonant contribution to  the 
$\Lambda_c(2595)$ and  $\Lambda_c(2625)$ decays into $\Lambda_c^+$ and two pions are
\begin{equation}
 \Gamma[\Lambda_c(2595)\to \Lambda_c \pi\pi ] = 1.84 \times g^2_{\Lambda_c(2595)\Sigma_c\pi}~{\rm [MeV]}\, , \qquad \Gamma[\Lambda_c(2625)\to \Lambda_c \pi\pi ] = 0.27 \times g^2_{\Lambda_c(2625)\Sigma_c^*\pi}~{\rm [MeV]} \label{eq:resulwdth}
\end{equation}
with the $\pi^0\pi^0$ channel being the 81.5\% and 45.0\% of the total for the $\Lambda_c(2595)$ and $\Lambda_c(2625)$ partial widths, respectively. In the exact isospin limit, the two-neutral-pions   partial width is a factor of two smaller than the $\pi^+\pi^-$ one.   The experimental 
width of the $\Lambda_c(2595)$ is $2.6 \pm 0.6$ MeV (nearly 100\% saturated by the two $\Lambda_c\pi\pi$ modes), while there exists an upper bound of 0.97 MeV for   the $\Lambda_c(2625)$~\cite{Tanabashi:2018oca}. Hence, the experimental $\Gamma[\Lambda_c(2595)]$ provides a direct measurement of the $S-$wave coupling constant $g^2_{\Lambda_c(2595)\Sigma_c\pi}$, assuming that a possible $D-$wave contribution is negligible~\cite{Aaltonen:2011sf}.  The bound on $\Gamma[\Lambda_c(2625)]$, on the other hand, puts upper limits on the coupling in $S-$wave of this resonance to the $\Sigma_c^*\pi$ pair, but one should bear in mind that in this case, the resonant contribution does not saturate the decay width.

The interference term in Eq.~(\ref{eq:chwdth}) for the charged mode, and the equivalent one in the case of $\pi^0\pi^0$, gives a small contribution to the integrated width.  In particular for the $\Lambda_c(2595)$, it is  of the order of $-1\%$ and $-0.2\%$ for the $\pi^+\pi^-$
and $\pi^0\pi^0$ channels, respectively. For the $\Lambda_c(2625)$, it becomes larger around  $-5\%$ and $-4\%$, respectively, but it is still quite small. This can be easily understood by changing the $s_{12}$ and $s_{23}$ integration variables to $E_3$ and $\cos\theta_{23}$, with $\theta_{23}$ the angle formed by the two pions 
in the resonance rest-frame.  The  energy $E_2$ (or equivalently $s_{13}$) depends on $\cos\theta_{23}$ through the conservation of energy equation, $M_R=E_3(\vec{p}_3\,)+E_2(\vec{p}_2)+E_{\Lambda_c}(\vec{p}_2+\vec{p}_3)$. In the infinite charm limit, the recoiling $\Lambda_c$ baryon carries off momentum but not kinetic energy, and hence the approximation~\cite{Cho:1994vg}
\begin{equation}
 E_2 \sim  M_R-M_{\Lambda_c}-E_3 \label{eq:approx}
\end{equation}
turns out to be quite accurate, specially for the $\Lambda_c(2595)$ where the energy released by by the decaying resonance is very small. Within this approximation, the only dependence of the differential decay rate on $\cos\theta_{23}$  comes from the scalar product $\vec{p}_2\cdot\vec{p}_3$ in the interference term, that would vanish in the integrated width, since $\cos\theta_{23}$ covers almost totally the $[-1,1]$ range for all $E_3$ allowed values. Indeed, we recover 
Eq.~(3.5), up to a factor 1/2, of Ref.~\cite{Cho:1994vg} from the expression of Eq.~(\ref{eq:chwdth}) by neglecting the interference term and adopting 
the approximation of Eq.~\eqref{eq:approx}, using that $ds_{12}=2M_RdE_3$ and taking into account that the integration over $ds_{23}$ gives $4p^*_{\pi^+}p^*_{\pi^-}\sim 4\sqrt{E_3^2-m_\pi^2}\sqrt{E_2^2-m_\pi^2}$, approximating in the propagators $(s_{12(13)}-M^2_{\Sigma_c^{(*)}})$ by $2M_{\Sigma_c^{(*)}} (M_R-E_{3(2)}-M_{\Sigma_c^{(*)}})$, and finally identifying $g_D^2= h_1^2/2$ and $g^2_{R\Sigma_c^{(*)}\pi}= 3h_2^2(M_R-M_{\Sigma_c^{(*)}})^2/(2f_\pi^2)$, with $h_{1,2}$ used in Ref.~\cite{Cho:1994vg}.  The factor 1/2 introduced in this latter work does not hold for the $\pi^+\pi^-$ decay mode, though should be included for the neutral mode  
\footnote{Note, however, that the expression for the $\Lambda_c\pi^0\pi^0$ partial width used by the CDF Collaboration in Ref.~\cite{Aaltonen:2011sf} is wrong by a factor of two. The 1/2 in Eq.~(13) for the amplitude in that reference should be replaced by $1/\sqrt{2}$.}.

Finally, we would like to mention that three-body $\Lambda_c(2595)\to \Lambda_c\pi\pi$ decay rate  can be approximated by using the narrow width approximation of the $\Sigma_c^{(*)}-$propagators,
\begin{equation}
 \left|G_{\Sigma_c}(s)\right|^2 \sim \frac{\pi}{M_{\Sigma_c} \Gamma_{\Sigma_c}(s)} \delta(s-M^2_{\Sigma_c}) = \frac{6\pi f_\pi^2}{M_{\Lambda_c} |\vec{p}_\pi^{\,\Lambda_c\pi}|^3} \delta(s-M^2_{\Sigma_c}) \label{eq:delta-appx}
\end{equation}
which leads to 
\begin{eqnarray}
 \Gamma_{\pi^-\pi^+} &\sim& ( \Gamma_{\Lambda_c(2595)\to \Sigma_c^{++}\pi^-} + \Gamma_{\Lambda_c(2595)\to \Sigma_c^{0}\pi^+}) \, , \qquad \Gamma_{\pi^0\pi^0} \sim \Gamma_{\Lambda_c(2595)\to \Sigma_c^{+}\pi^0}\label{eq:2body}\\
 \Gamma_{\Lambda_c(2595)\to \Sigma_c^a\pi^b}  &=& \frac{g^2_{\Lambda_c(2595)\Sigma_c\pi}}{6\pi} \frac{M_{\Sigma_c^a}}{M_{\Lambda_c(2595)}}|\vec{p}_\pi|\, ,\quad |\vec{p}_\pi|= \frac{\lambda^{\frac12}(M_{\Lambda_c(2595)}^2,M_{\Sigma_c^a}^2,m_{\pi^b}^2)}{2M_{\Lambda_c(2595)}}
\end{eqnarray}
for the charge-combinations $(a,b)= (++,-),(0,+)$ and $(+,0)$, which correspond to the square of the amplitudes of each of the three diagrams depicted in Fig.~\ref{fig:decay}. To obtain Eq.~\eqref{eq:2body} from Eq.~\eqref{eq:chwdth}, using the approximation of Eq.~\eqref{eq:delta-appx}, we have neglected the  interference contributions,  have approximated the LAB energy  of the non-resonant pion and  the momentum of the resonant one  by $(M_{\Lambda_c(2595)}-M_{\Sigma_c})$ and $|\vec{p}_\pi|$, respectively, and in addition,  we have made use that the momentum of the non-resonant pion in the $\Sigma_c-$rest frame  is $M_{\Lambda_c(2595)}|\vec{p}_\pi|/M_{\Sigma_c}$. The two body $S-$wave-widths limit of Eq.~\eqref{eq:2body} works well when the intermediate $\Sigma_c$ is nearly on-shell. The value used here for $M_{\Lambda_c(2595)}$ is 1.2 (4.6) MeV  below (above) the $\Sigma_c^{++,0}m_{\pi^\mp}$ ($\Sigma_c^{+}m_{\pi^0}$) threshold.  We find that $\Gamma_{\pi^0\pi^0}$ and  $\Gamma_{\Lambda_c(2595)\to \Sigma_c^{+}\pi^0}$ differ only in 0.19 $g^2_{\Lambda_c(2595)\Sigma_c\pi}$ [MeV], this is to say, the latter width is  just 2.5\% greater than the former one. The $\Sigma_c^{++}$ and $\Sigma_c^{0}$ cannot be put on shell for this mass of the $\Lambda_c(2595)$, but clearly the differential decay width of Eq.~\eqref{eq:chwdth} is strongly dominated by the contribution of two well separated peaks that correspond to the first two mechanisms shown in Fig.~\ref{fig:decay}~\cite{Cho:1994vg}.  
\section{Results and discussion}
\label{sec:res}
\subsection{${\rm SU(6)}_{\rm lsf}\times$HQSS hadron molecules: dependence on the renormalization scheme}
\label{sec:Su6results}
\begin{figure}[h]
\begin{center}
\makebox[0pt]{\includegraphics[width=0.4\textwidth]{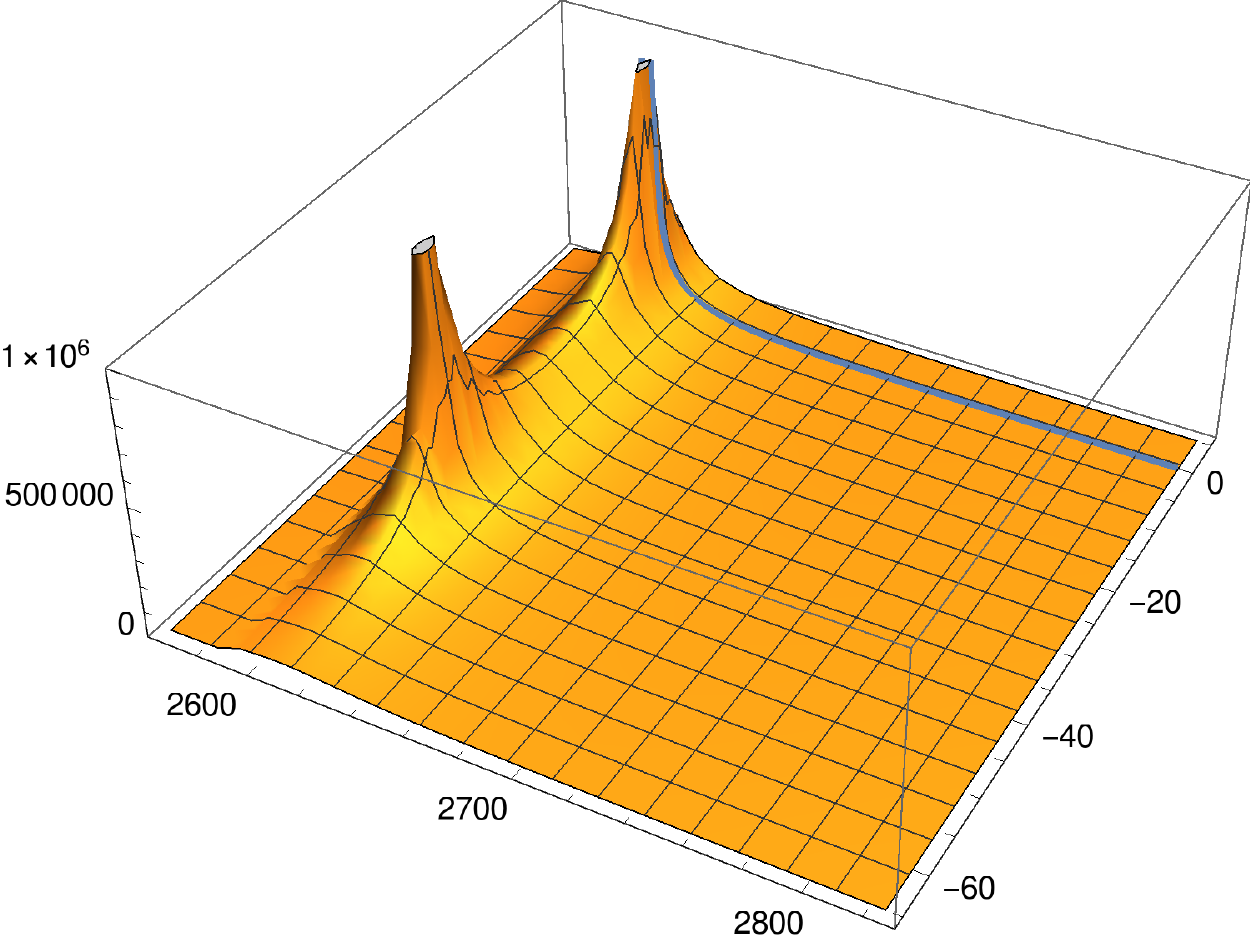}\hspace{0.75cm}\includegraphics[width=0.4\textwidth]{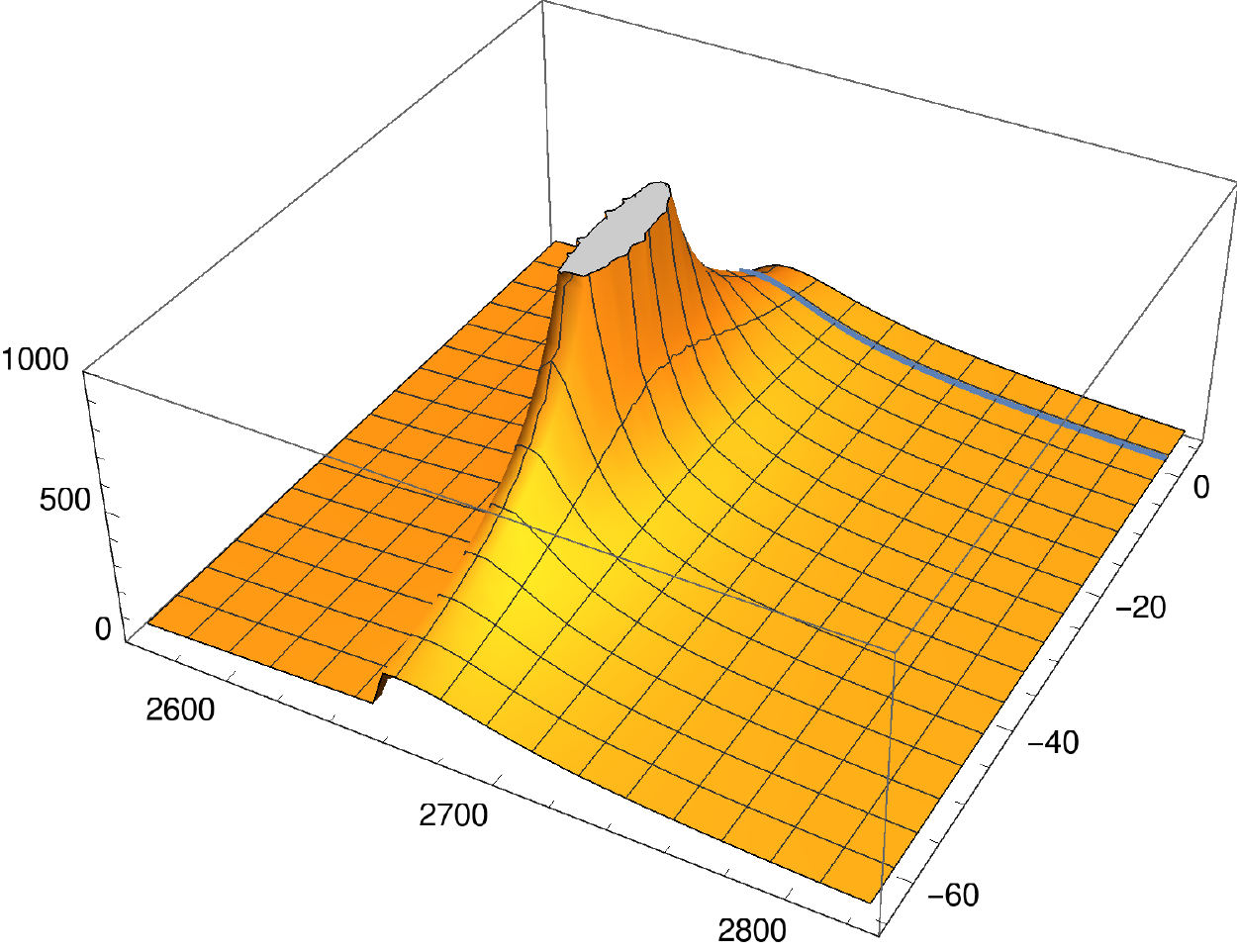}}\\\vspace{0.5cm}
\makebox[0pt]{\includegraphics[width=0.4\textwidth]{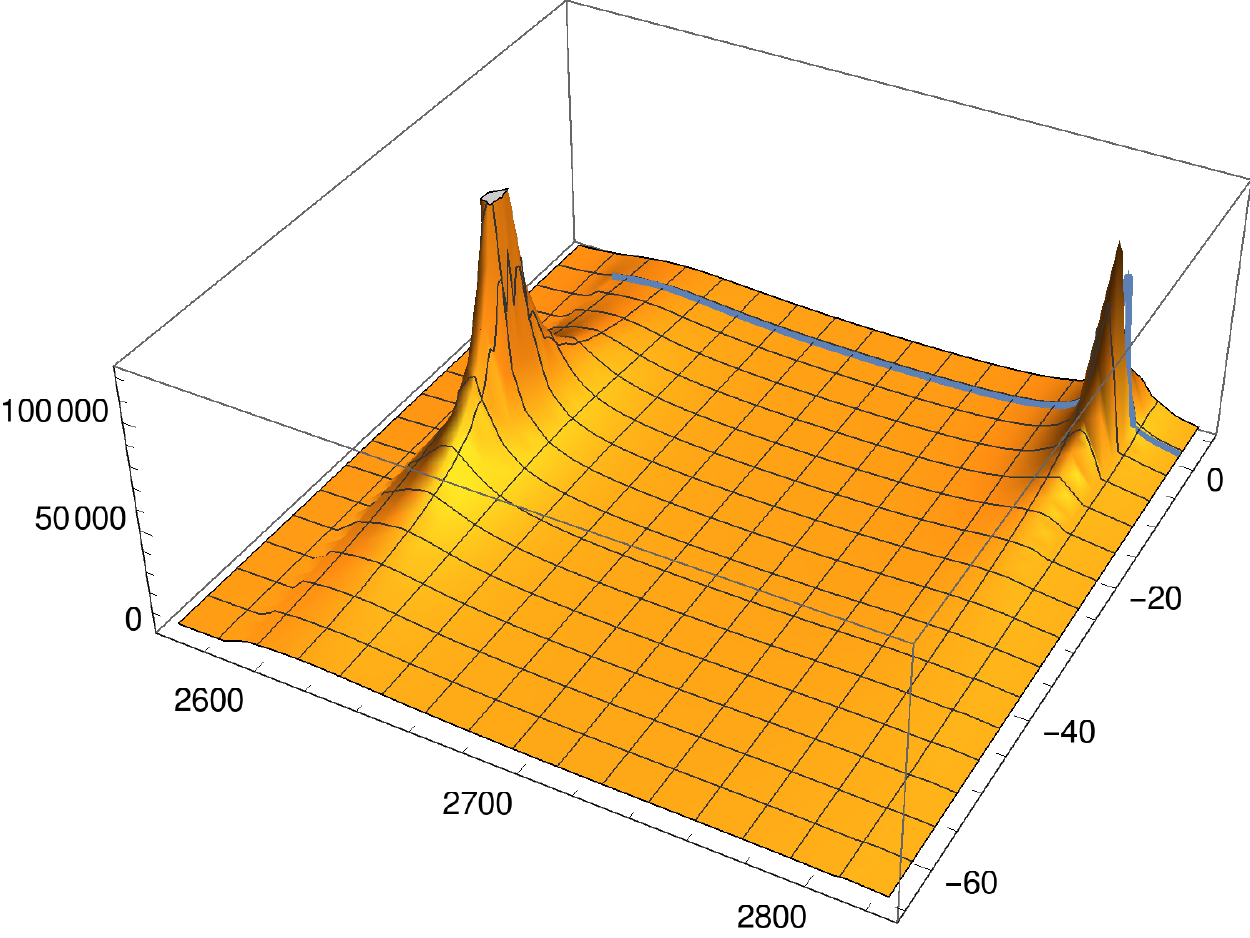}\hspace{0.75cm}\includegraphics[width=0.4\textwidth]{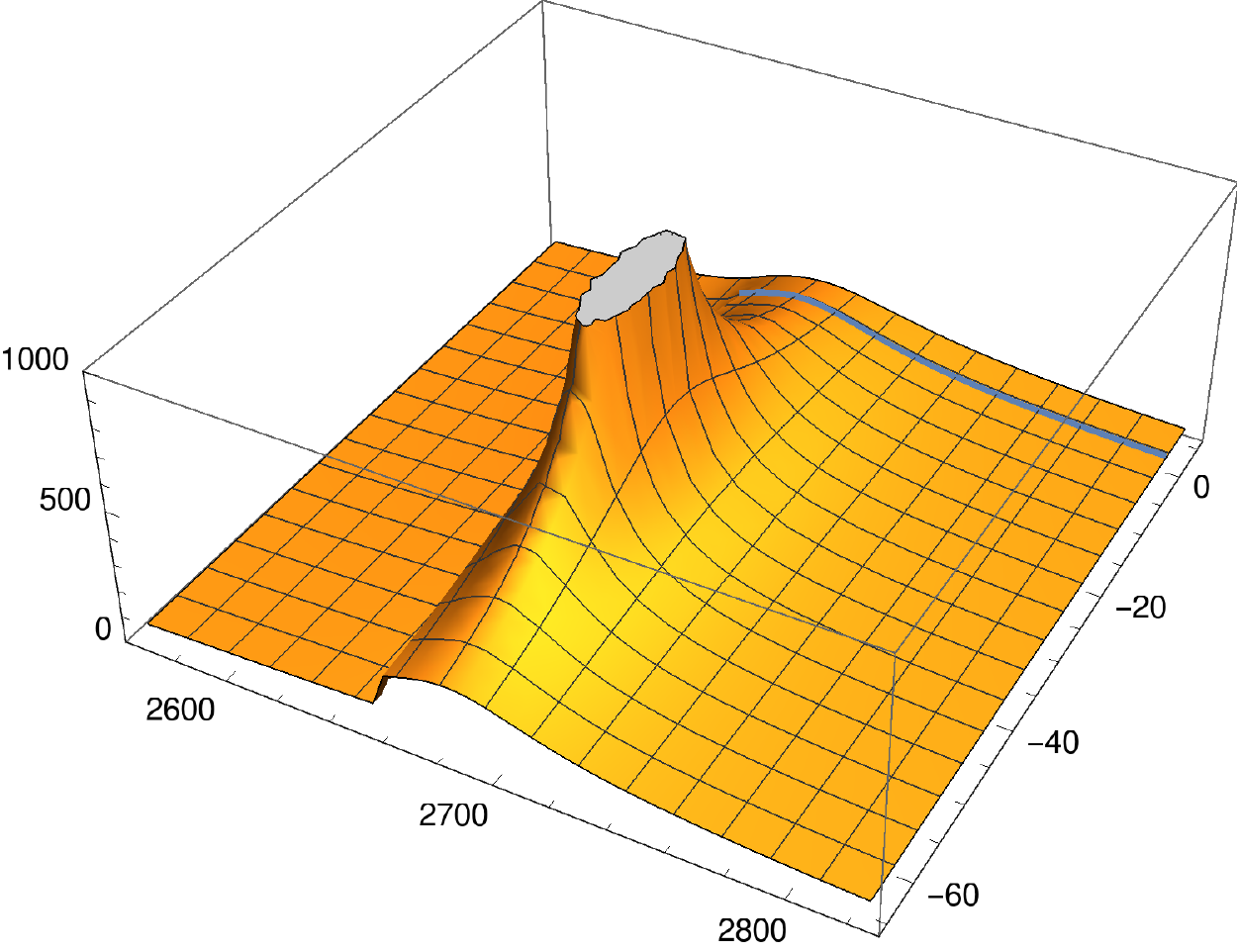}}
\end{center}
\caption{Absolute value of the determinant  of the $T-$matrix (AVD-$T$) in the $J^P=1/2^-$ (left) and $J^P=3/2^-$ (right) sectors using  
two UV renormalization schemes: SC$\mu$ with $\alpha=0.95$  and a cutoff of 650 MeV  in the top and  bottom panels respectively. We display the  AVD-$T$ for both the FRS (${\rm Im}(E)> 0$) and the SRS (${\rm
    Im}(E)< 0$) [fm] of the unitarized amplitudes as a function
  of the complex energy $E$ [MeV]. We also show the scattering line (blue solid curve) in all the cases. Bare CQM exchange interactions are set to zero 
  ($d_1=c_1=0$ in Eq.~\eqref{eq:exchange}). In the $z-$axis we have AVD-$T$, in the $x-$axis we have ${\rm Re}(E)$ and
in the $y-$axis we have ${\rm Im}(E)$.}
  \label{fig:noCQM}
\end{figure}
First we present in Fig.~\ref{fig:noCQM} the dynamically generated resonances (poles in the SRS of the amplitudes) that  
are obtained, when the effects produced by the exchange of CQM bare states are neglected. We show both the $J=1/2$ and $J=3/2$ sectors, and consider the two renormalization schemes introduced in Subsec.~\ref{sec:renor}. The numerical positions of the poles and residues are given in the first row of Tables~\ref{tab:j32} and \ref{tab:j12}. 
\begin{table}
 \begin{tabular}{cc|cccc|cccc}
            &            & \multicolumn{4}{c|}{$\Lambda=650$ MeV} & \multicolumn{4}{c}{SC$\mu$ ($\alpha=0.95$)} \\
 $ d_1$ & $ c_1$ & $M-i\,\Gamma/2$ & Type & $|g_{\Sigma_c^*\pi}|$ & $|g_{ND^*}|$ & $M-i\,\Gamma/2$ & Type & $|g_{\Sigma_c^*\pi}|$  & $|g_{ND^*}|$\\\hline
 0     &  0 & $(2680.4 -i\,33.0)$ & $1^- $ & 2.0 & 2.5& $(2662.6 -i\,27.2)$ &  $1^- $ & 2.3 & 2.4  \\\hline
 $-0.8$  &  0 & $(2704.0 -i\,31.5)$ & $1^- $ & 1.7 &  2.2 & $(2688.8 -i\,28.4)$ & $1^- $ &  1.8 &    1.9\\
 $-0.8$  &  0 & 2615.8 & CQM &  1.0  & 0.6 & 2617.5 & CQM  & 1.1 & 0.5\\\hline
 $-0.8$  &  $-1$ & $(2706.8 -i\,30.2)$ & $1^- $ & 1.7 &  2.6 & $(2681.1-i\,27.0)$ & $1^- $ &1.8  &2.4   \\
 $-0.8$  &  $-1$ &  2614.9 & CQM  & 1.1  & 0.2 & 	2617.7 & CQM & 1.0 & 0.5\\\hline
 $-0.8$  &  1 & $(2701.4 -i\,32.4)$ & $1^- $ & 1.8 &  1.8 & $(2695.5 -i\,30.6)$  & $1^- $ & 1.8 & 1.5   \\
 $-0.8$  &  1 & 2620.9 & CQM & 0.9  & 1.5 &  2612.9 & CQM & 1.2 & 1.4\\\hline
 \end{tabular}
\caption{Properties of the CQM and molecular $J^P=3/2^-$ poles for different renormalization schemes and  values of the $d_1$ and $c_1$ LECs, which determine the interplay between CQM and baryon-meson pair degrees of freedom. The angular momentum--parity quantum-numbers of the {\it ldof} are always $1^-$, and masses and widths are given in MeV units.}
\label{tab:j32}
\end{table}
\begin{table}
 \begin{tabular}{cc|ccccc|ccccc}
            &            & \multicolumn{5}{c|}{$\Lambda=650$ MeV} & \multicolumn{5}{c}{SC$\mu$ ($\alpha=0.95$)} \\
 $d_1$ & $ c_1$ & $M-i\,\Gamma/2$ & Type & $|g_{\Sigma_c\pi}|$ &  $|g_{ND}|$ & $|g_{ND^*}|$ & $M-i\,\Gamma/2$ & Type  & $|g_{\Sigma_c\pi}|$  & $|g_{ND}|$ & $|g_{ND^*}|$\\\hline
 0     &  0 & $(2609.9 -i\,28.8)$ & $1^-$  & 2.0 &  2.3 & 0.7& $(2608.9 -i\,38.6)$ & $1^-$ &   2.3 & 2.0 & 1.9   \\
  0    &  0  & $(2798.7 -i\,2.0)$  & $0^-$  & 0.3 & 1.8  &  4.1 & $(2610.2 -i\, 1.2)$ & $0^-$ &  0.5 &  3.9 & 6.2 \\\hline
 $-0.8$  &  0 & 2590.0 & $1^-$ & 1.1 &  1.0 &  0.3 & 2591.1$^*$ & $1^-$ & 1.3 &   1.5 &    0.5\\
 $-0.8$  &  0 & $(2799.0 -i\,2.2) $  & $0^-$  & 0.3  & 1.8&   4.1& $(2611.6 - i\,0.4)$  & $0^-$ & 0.3 &   3.5 &   6.2 \\
 $-0.8$  &  0 & $(2659.1 -i\,17.3)$  & CQM  & 1.3  &  1.6 & 0.3 & $(2652.8 -i\,22.2 )$ &  CQM & 1.5 &   0.8  & 1.7\\\hline
 $-0.8$  &  $-1$ & 2589.3 & $1^-$ & 1.2 &  0.5 &  0.2 & $(2589.4-i\,8,6)^*$ & $1^-$ & 2.6 &   1.7 &    0.7\\
 $-0.8$  &  $-1$ & $(2800.7 -i\,2.3) $  & $0^-$  & 0.3  & 1.7&   4.1&  $(2610.1 -i\,0.0 )$ & $0^-$ & 0.0 &   3.8 &   5.9 \\
 $-0.8$  &  $-1$ & $(2657.2 -i\,15.8)$  & CQM  & 1.2  &  2.3 & 0.5 & $(2642.3 - i\,19.4)$ &  CQM & 1.5 &   1.1  & 2.5\\\hline
 $-0.8$  &  1 & 2591.0 & $1^-$ & 0.9 &  1.2 &  0.4 & 2590.3 & $1^-$ & 1.1 &   1.7 &    0.5\\
 $-0.8$  &  1 & $(2798.8 -i\,1.9) $  & $0^-$  & 0.3  & 1.8&   4.1& $(2612.6 - i\,0.7)$  & $0^-$ & 0.4 &   3.2 &   6.3 \\
 $-0.8$  &  1 & $(2660.0 -i\,18.9)$  & CQM  & 1.3  &  1.1 & 0.2 & $(2659.6 -i\,26.8 )$ &  CQM & 1.6 &   0.7  & 1.2\\\hline
 \end{tabular}
\caption{Properties of the CQM and molecular $J^P=1/2^-$ poles for different renormalization schemes  and values of the $d_1$ and $c_1$ LECs, which determine the interplay between CQM and baryon-meson pair degrees of freedom. Molecular states are labeled according to their dominant {\it ldof} configuration, $0^-$ or $1^-$,  and masses and widths are given in MeV units. $^*$: Virtual state placed in the 100 sheet below the $\Sigma_c\pi$ threshold.}
\label{tab:j12}
\end{table}

The SC$\mu-$results found here, working with the reduced $ND^{(*)}-\Sigma_c^{(*)}\pi$ coupled-channels space, reproduce reasonably well the most important features reported in the original works of Refs.~\cite{GarciaRecio:2008dp, Romanets:2012hm}. Indeed, we choose $\alpha=0.95$ to better account for some of the effects produced by the channels that have not been considered in the current approach. We see that a narrow $J^P=1/2^-$ $\Lambda_{c\, (n)}^{\rm 0^-}(2595)$ resonance ($\Gamma \sim 2$ MeV) is produced. This is  mostly generated from the extended WT $ND-ND^*$ coupled-channels dynamics in the $j_{ldof}^\pi=0^-$ subspace. This state has a small coupling to the $(j_{ldof}^\pi=1^-)$ $\Sigma_c\pi$ channel which, in addition to the proximity to the open threshold,  explains its small width.  There appears a second $J^P=1/2^-$ pole [$\Lambda_{c\, (b)}^{\rm 1^-}(2595)$]  in the 2.6 GeV region. Although it is  placed relatively close to the $\Sigma_c\pi$ threshold, this resonance is broad ($\Gamma \sim 75$ MeV) because of its sizable coupling to the latter open channel. Nevertheless, as seen in  Fig.~\ref{fig:noCQM}, this second wide state will not produce visible effects on the baryon-meson $S-$wave cross sections, since its possible impact for real values of $s$ will be shadowed by the narrow $\Lambda_{c\, (n)}^{\rm 0^-}$ that is located at a similar mass and much closer to the scattering line. Thus, this double pattern structure would be difficult to be confirmed experimentally, and it will not certainly show up in the $\Lambda_c\pi\pi$ spectrum, where the evidences of the $\Lambda_c(2595)$ have been reported~\cite{Edwards:1994ar, Albrecht:1997qa,Aaltonen:2011sf}. However, it has been argued that exclusive semileptonic $\Lambda_b$ ground-state decays into excited charmed  
$\Lambda_c^*$ baryons could unravel the two $\Lambda_c(2595)$ states~\cite{Liang:2016exm,Nieves:2019kdh}, if they exist.

In the $J^P=3/2^-$ sector, we find a  resonance that  clearly is the HQSS
partner of the broad  $J^P=1/2^-$ $\Lambda_{c\, (b)}^{\rm 1^-}(2595)$
state, with quantum numbers  $1^-$ for $j^\pi_{ldof}$. It is located above the $\Sigma^*_c\pi$ threshold, with a width of around 55 MeV.  Furthermore, 
the coupling of this $J^P=3/2^-$ pole to the $\Sigma_c^*\pi$ channel is essentially identical to that of the $\Lambda_{c\, (b)}^{\rm 1^-}(2595)$ to $\Sigma_c\pi$. This $J^P=3/2^-$ isoscalar resonance might be identified  with the $D-$wave $\Lambda_c(2625)$, although its mass and width significantly differ from those of the physical state. In Refs.~\cite{GarciaRecio:2008dp} and \cite{Romanets:2012hm}, it is argued that 
a change in the renormalization subtraction constant  could  move the resonance down by 40 MeV to the nominal position of the physical state, and that in addition, this change of the mass  would considerably reduce the width, since the state might even become  bound  below the
$\Sigma_c^*\pi$ threshold. Thus, within the ${\rm SU(6)}_{\rm lsf} \times {\rm SU(2)}_{\rm HQSS}$ model, the $\Lambda_c(2625)$ would turn out to be the HQSS partner of the second broad 
$\Lambda_{c\, (b)}^{\rm 1^-}(2595)$ pole instead of the narrow  $\Lambda_{c\, (n)}^{\rm 0^-}(2595)$ resonance\footnote{A more detailed discussion, incorporating some elements of group theory, can be found in \cite{Romanets:2012hm} and in Subsection 2.2.1 of Ref.~\cite{Nieves:2019kdh}.}. This is in sharp contrast to  the predictions of the CQMs,  where  there is no a second 2595 pole, and  the $\Lambda_c(2625)$ and the narrow $\Lambda_c(2595)$ are HQSS siblings, produced by a $\lambda-$mode excitation of the ground $1/2^+$ $\Lambda_c$ baryon.

The SC$\mu-$renormalization scheme plays an important role in enhancing the influence of the $N D^*$ channel in the dynamics of the narrow $\Lambda_{c\, (n)}^{\rm 0^-}(2595)$ state. Indeed, this scheme also produces a reduction in the mass of the resonance of around 200 MeV, which thus appears in the region of 2.6 GeV, instead of in the vicinity of the $ND$ threshold. Indeed, we see also in Fig.~\ref{fig:noCQM} and Table~\ref{tab:j12} that if the UV behaviour of the amplitudes is renormalized by means of a common momentum cutoff of 650 MeV (Eq.~\eqref{eq:uvcut}), the position of the $j_{ldof}^\pi=0^-$ pole in the $J^P=1/2^-$ sector moves up drastically, and it now appears at 2.8 GeV with little chances to be identified with the physical $\Lambda_c(2595)$ state. It is still narrow, because HQSS prevents its coupling to  $\Sigma_c\pi$ to become large. However, the main features of the broad $J^P=3/2^-$ resonance and the $j_{ldof}^\pi=1^-$ one in the $J=1/2$ sector are not much affected by the change of renormalization scheme. The mass position of the latter resonance can be moved down to the vicinity of the $\Sigma_c\pi$ threshold using cutoffs of the order of 750 MeV, still reasonable. At the same time its width also decreases since the available phase space for the decay becomes smaller. However, to obtain masses for the $3/2-$state of around 2625 MeV, significantly larger cutoffs of the order of 1200 MeV are needed. This might hint the existence of some  further contributions to those induced for the baryon-meson unitarity loops, and that are effectively accounted for the somehow unnatural  UV regulator. In this context, we will discuss in the next subsection effects produced by CQM degrees of freedom. In addition, the coupling $|g_{\Sigma_c^*\pi}|$ would take values of around 1.6 leading to $\Gamma[\Lambda_c(2625)\to \Lambda_c \pi\pi ]\sim 0.7$ MeV from Eq.~\eqref{eq:resulwdth}, 30\% below the upper bound on the total width of the resonance. However, taking into account that the $\Sigma_c^*-$resonant contribution measured by the ARGUS Collaboration is $(46\pm 14)\%$~\cite{Albrecht:1993pt} of the total, we find that 0.7 MeV is around two sigmas above the inferred upper bound for the resonant mechanism. Note that using Eq.~\eqref{eq:resulwdth},  the upper bound on the $\Sigma_c^*-$resonant contribution to the $\Lambda_c(2625)$ width leads to  
\begin{equation}
 |g_{\Sigma_c^*\pi}| < 1.3 \pm  0.2 \label{eq:upper}
\end{equation}

We end up this discussion by studying the relation between cutoff and SC$\mu$ UV renormalization schemes. Results obtained in SC$\mu$ are recovered by using appropriate channel-dependent cutoffs as detailed in Eq.~\eqref{eq:subtraction}. These are 459 MeV, 544 MeV, 905 MeV and 1044 MeV for $\pi \Sigma_c$, $\pi \Sigma_c^*$ $ND$ and $ND^*$, respectively. We see that the cutoff for  $N D^*$  is large and it enhances the importance of this channel in the dynamics of   the narrow  $\Lambda_{c\, (n)}^{\rm 0^-}(2595)$ resonance found in the SC$\mu-$scheme.

\subsection{CQM and baryon-meson degrees of freedom}
\label{sec:CQM-hadron}
\begin{figure}[h]
\begin{center}
\makebox[0pt]{\includegraphics[width=0.5\textwidth]{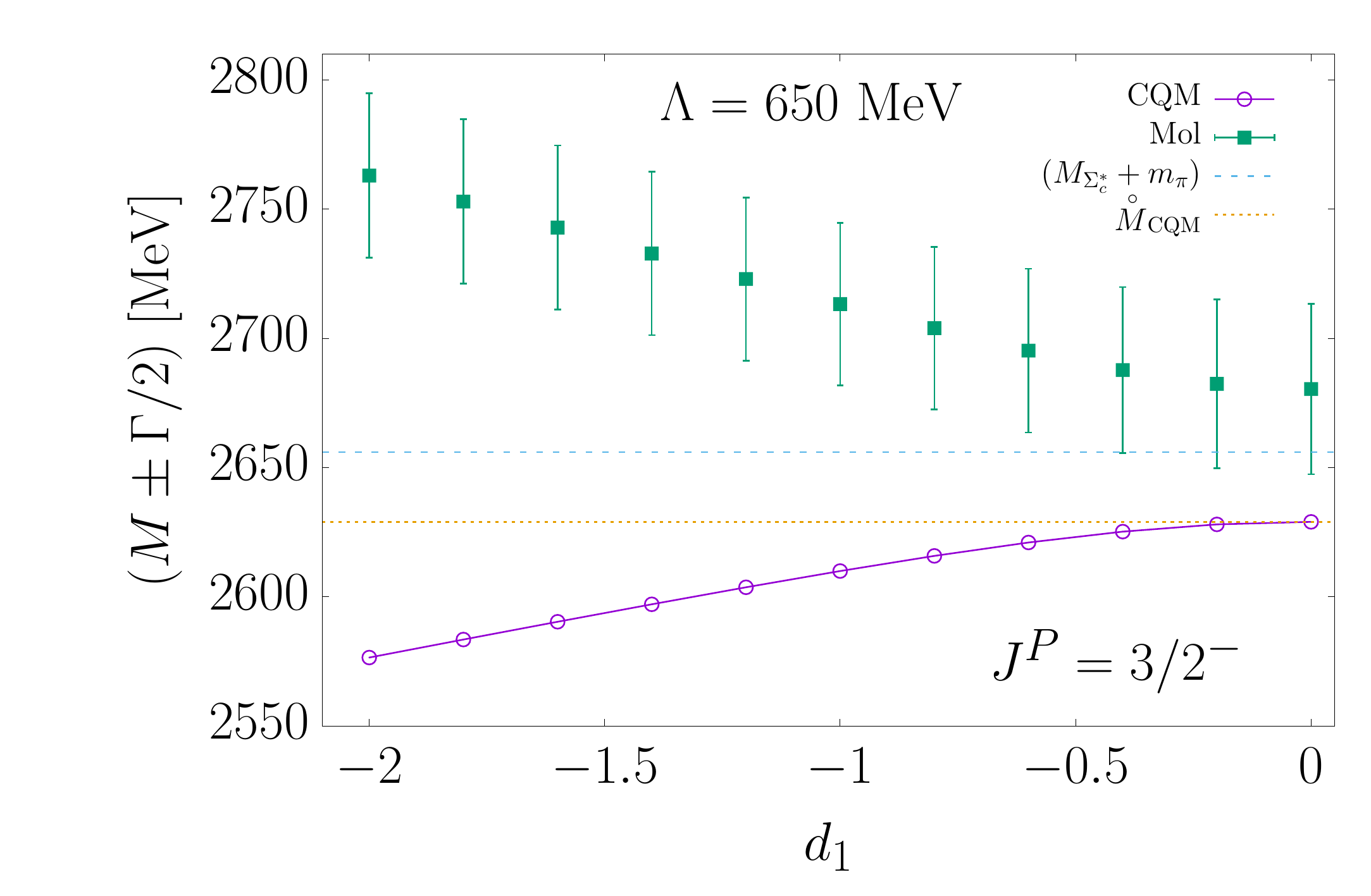}\hspace{0.25cm}\includegraphics[width=0.5\textwidth]{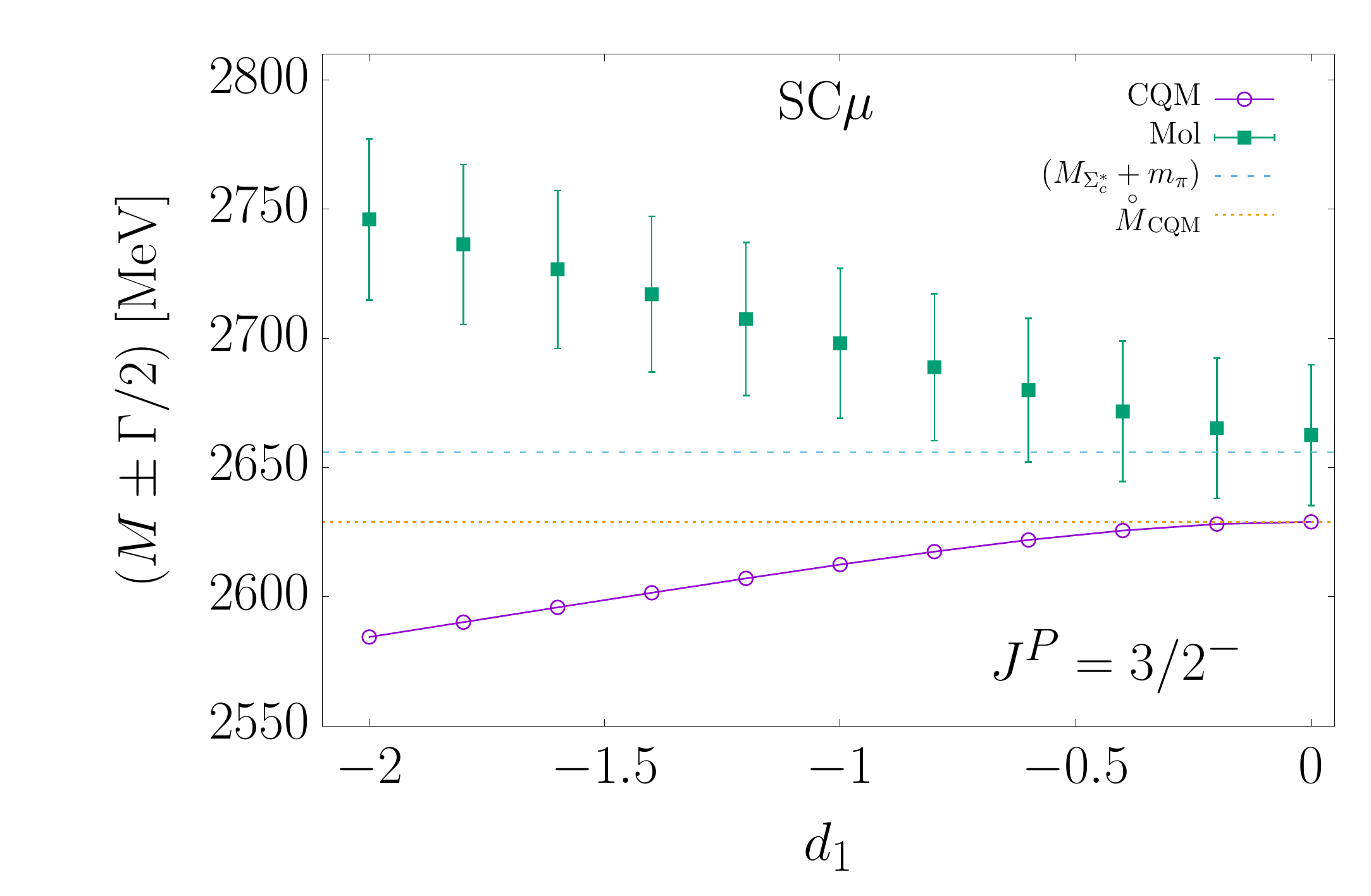}}
\end{center}
\caption{Dependence of the $J^P=3/2^-$ CQM and molecular pole positions as a function of the LEC $d_1$, for $c_1=0$. We show results for both, the cutoff and SC$\mu$ ($\alpha=0.95$) renormalization schemes, and the values of the bare CQM mass and the $\Sigma_c^*\pi$ threshold energy. \label{fig:32c1zero}}
\end{figure}
\begin{figure}[h]
\begin{center}
\makebox[0pt]{\includegraphics[width=0.4\textwidth]{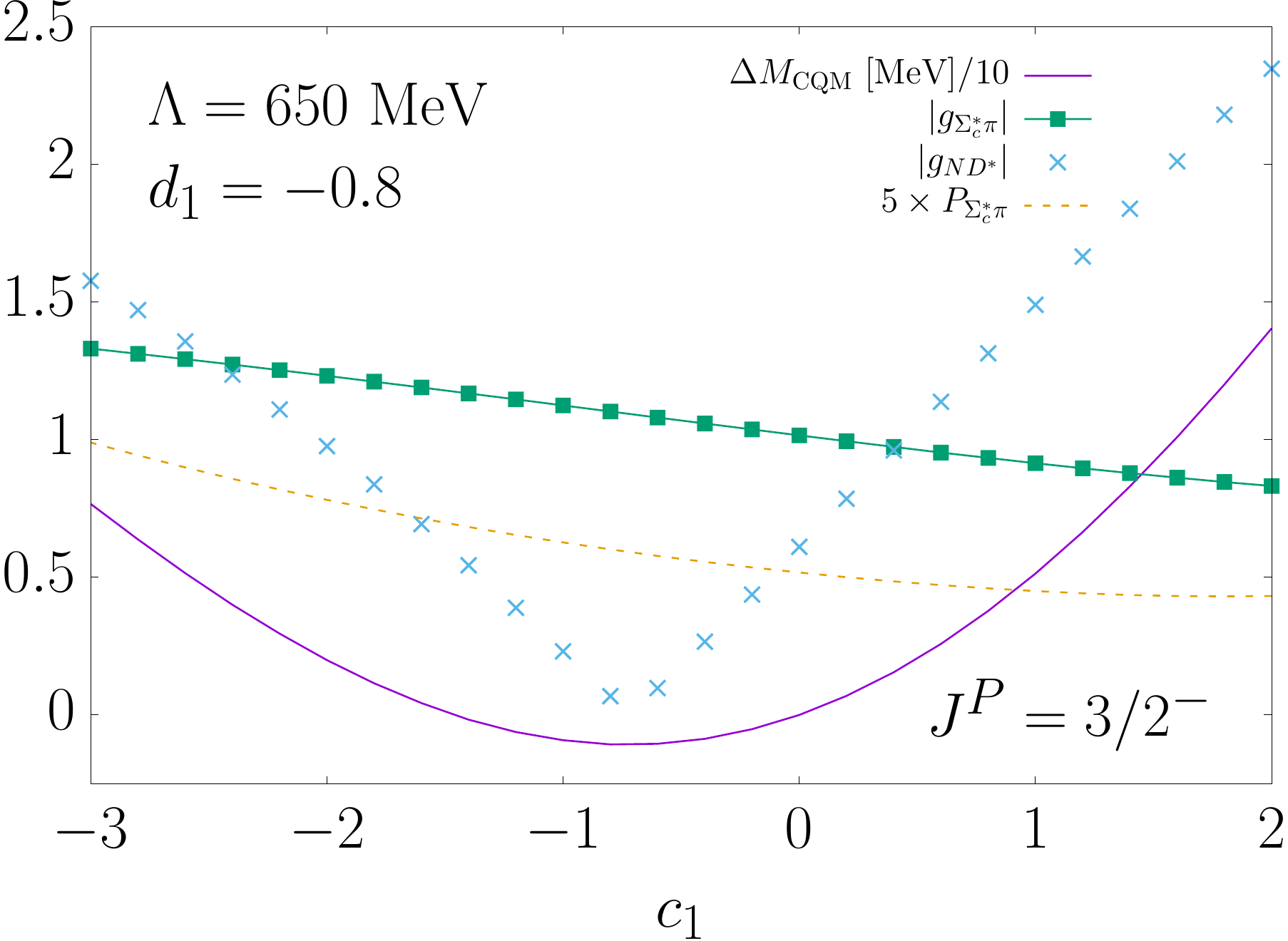}\hspace{0.75cm}\includegraphics[width=0.4\textwidth]{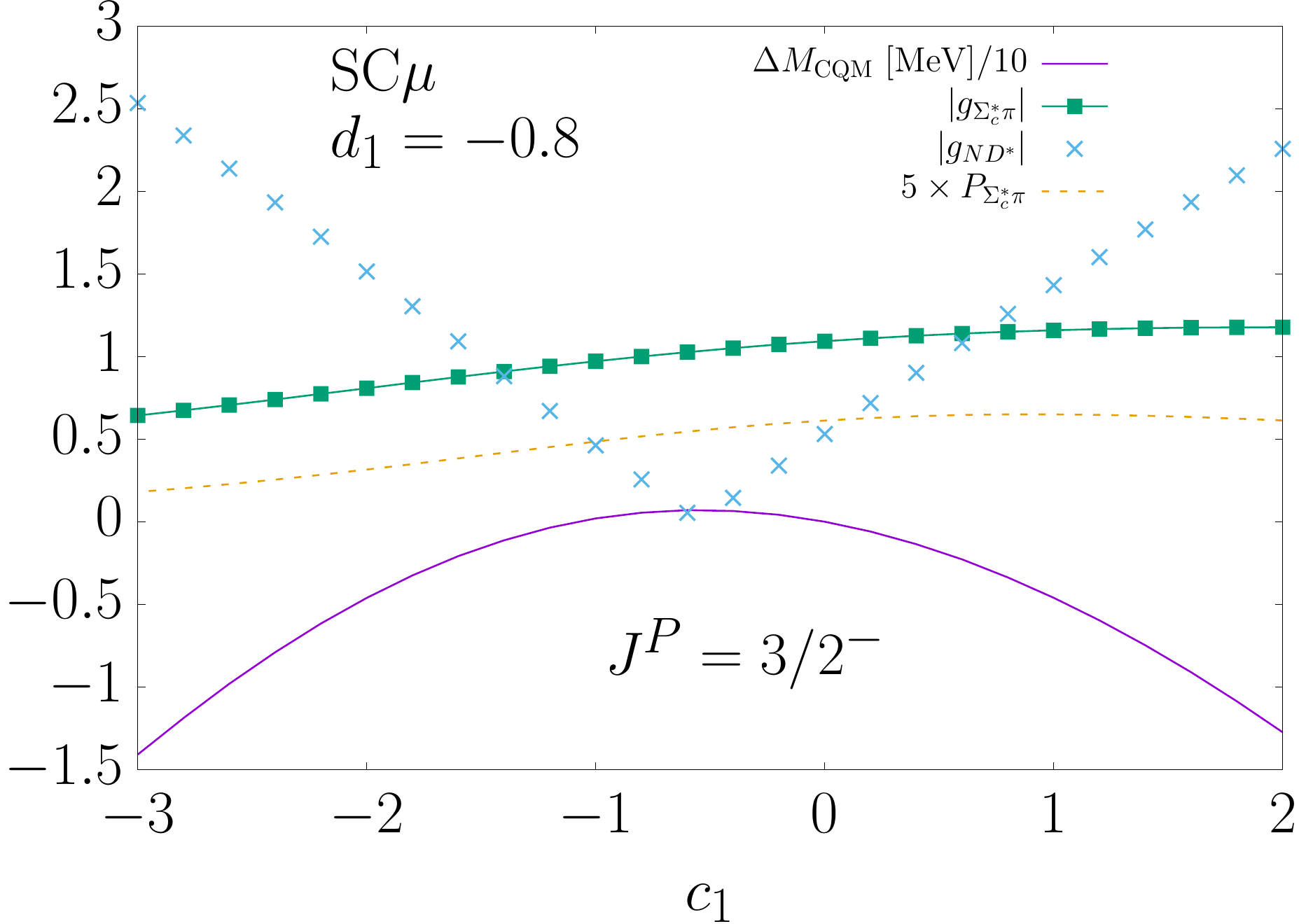}}
\end{center}
\caption{Dependence of the couplings, mass and $\Sigma^*_c \pi-$molecular probability of the $J^P=3/2^-$ dressed CQM pole as a function of the LEC $c_1$, for 
$d_1=-0.8$.  We show results for both, the cutoff and SC$\mu$ ($\alpha=0.95$) renormalization schemes. In addition, $\Delta M_{\rm CQM}= [M_{\rm CQM}(d_1)-M_{\rm CQM}(d_1=-0.8)]$, with  $M_{\rm CQM}(d_1=-0.8)= 2615.81$ MeV and 2617.49 MeV for the left and right panels, respectively.\label{fig:32c1nozero} }
\end{figure}

As mentioned in Subsec.~\ref{sec:QMth}, the quark model of Ref.~\cite{Yoshida:2015tia} predicts a $(J^P=1/2^-,3/2^-)$ HQSS doublet of states, almost degenerate and 
with  $\Mbare\sim 2629$ MeV. Though with some precautions, because the CQM bare mass is not an observable and  the matching procedure between the quark model and the effective hadron theory is not well defined, it seems natural to think that the bare CQM states should have a important influence in the dynamics of the $\Lambda_c(2595)$ and $\Lambda_c(2625)$ resonances, which are located so close. Indeed, the baryon-meson interactions of Eq.~\eqref{eq:exchange}, driven by the exchange of the CQM state, have a strong energy dependence close to $\Mbare\sim 2629$ that might be difficult to accommodate by just modifying the real part of the unitarity loops. The  LECs $c_1$ and $d_1$, that control the interplay between bare CQM and baryon-meson degrees of freedom, are unknown. They are also renormalization scheme-dependent,  and once the scheme is fixed, they should be inferred from data. The hope is that in this way, some theoretical predictions could become renormalization independent, at least in some energy window around the experimental inputs. 

\subsubsection{The $\Lambda_c(2625)$}

\begin{figure}[h]
\begin{center}
\makebox[0pt]{\includegraphics[width=0.26\textwidth]{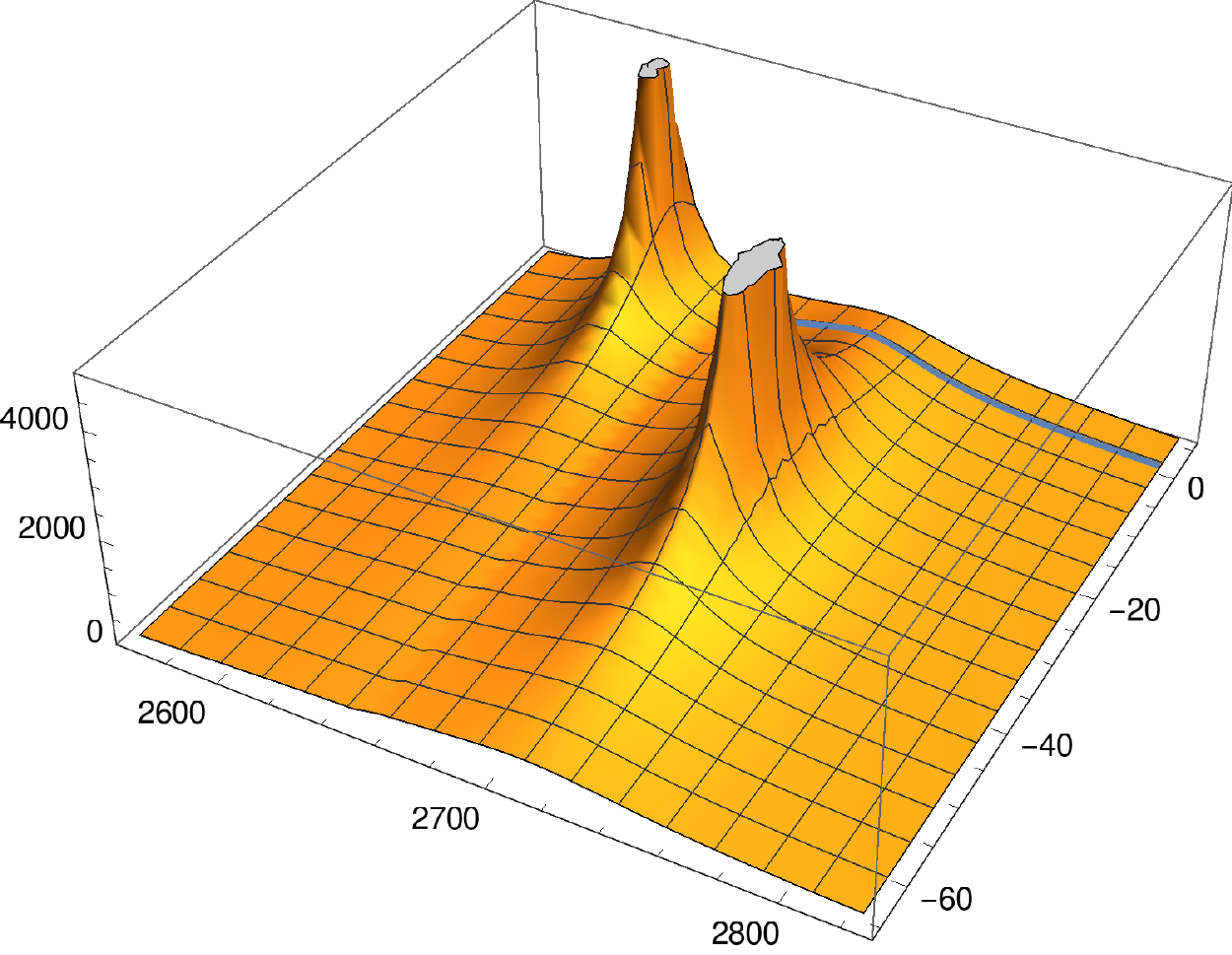}\hspace{0.75cm}\includegraphics[width=0.26\textwidth]{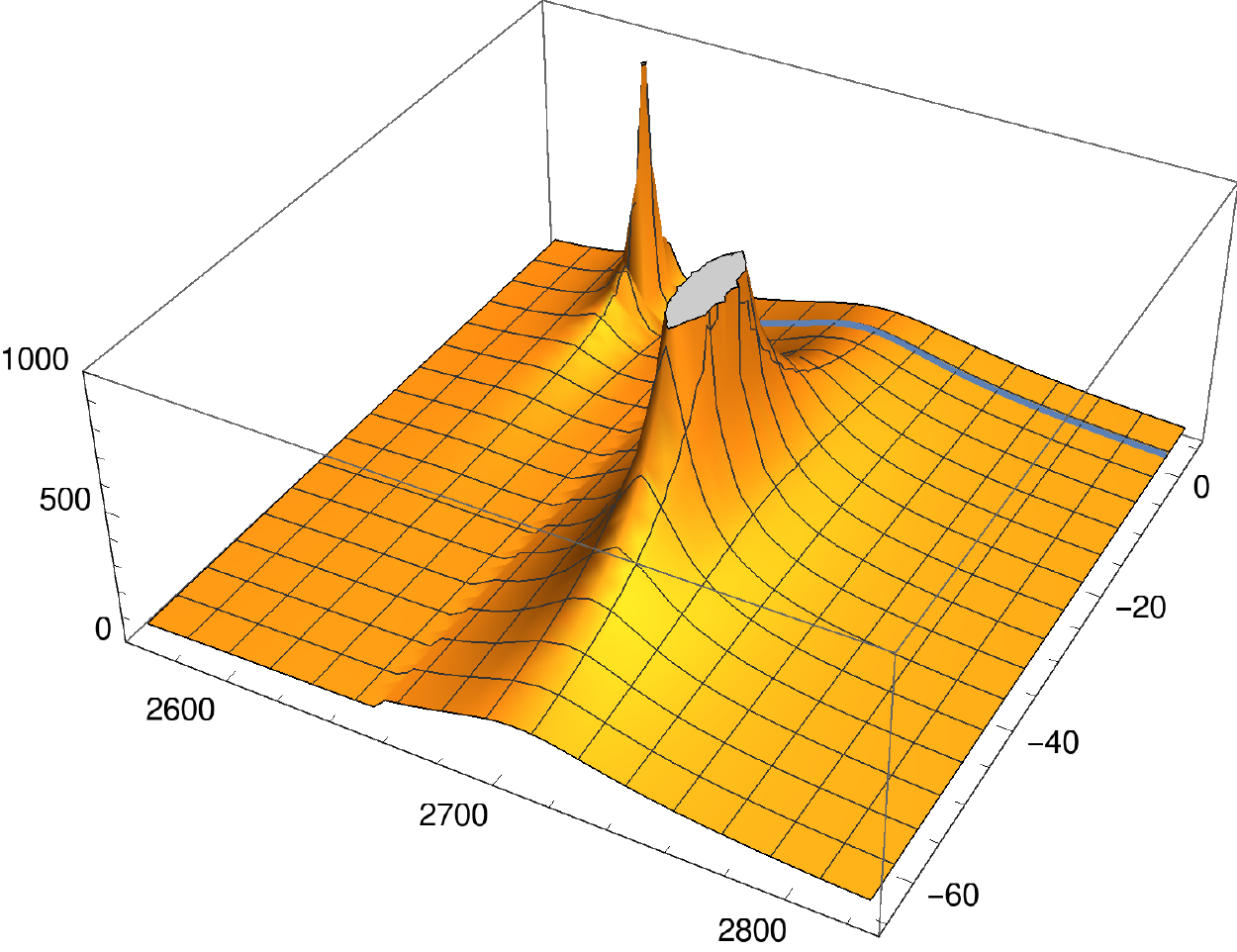}\hspace{0.75cm}\includegraphics[width=0.26\textwidth]{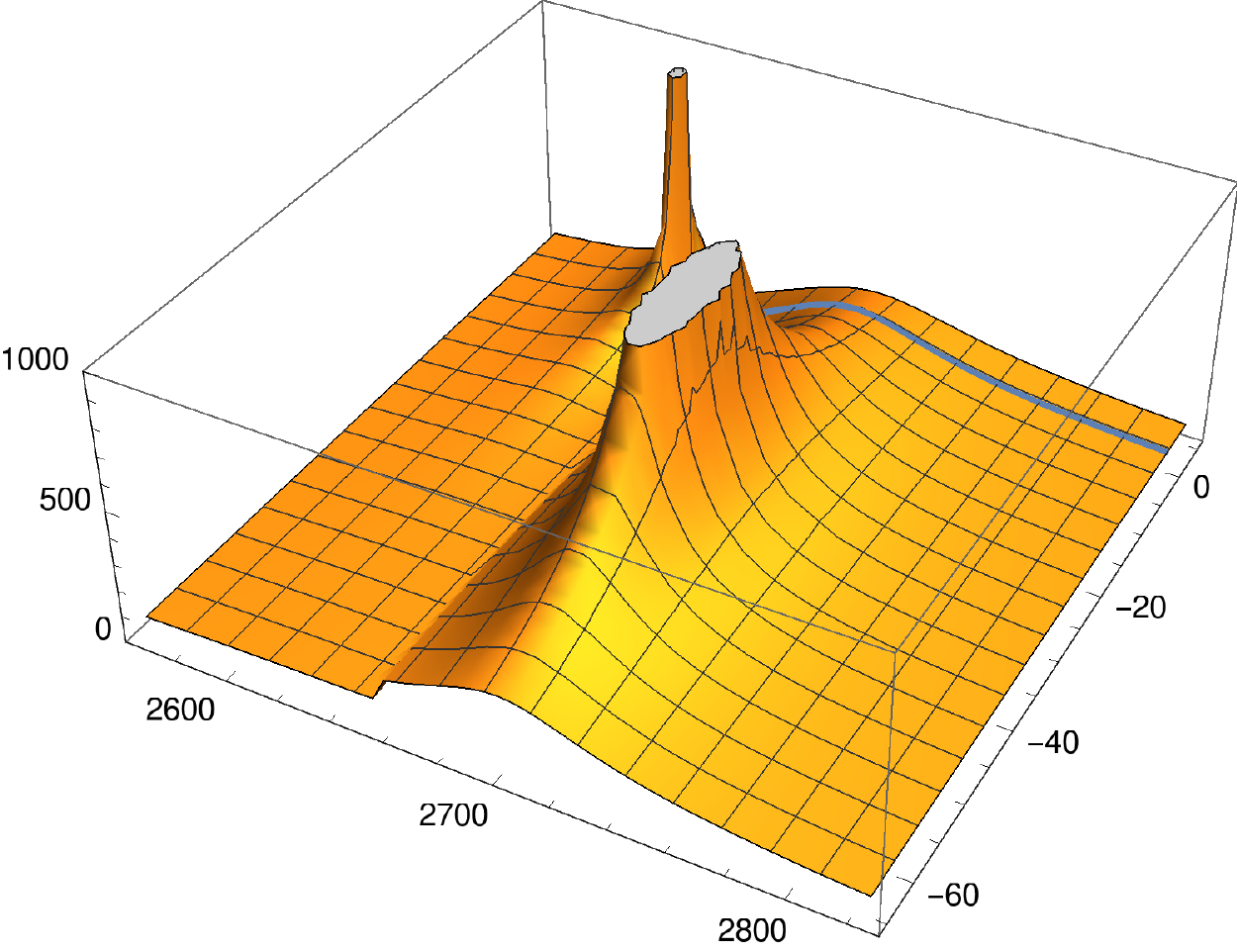}}\\\vspace{0.5cm}
\makebox[0pt]{\includegraphics[width=0.26\textwidth]{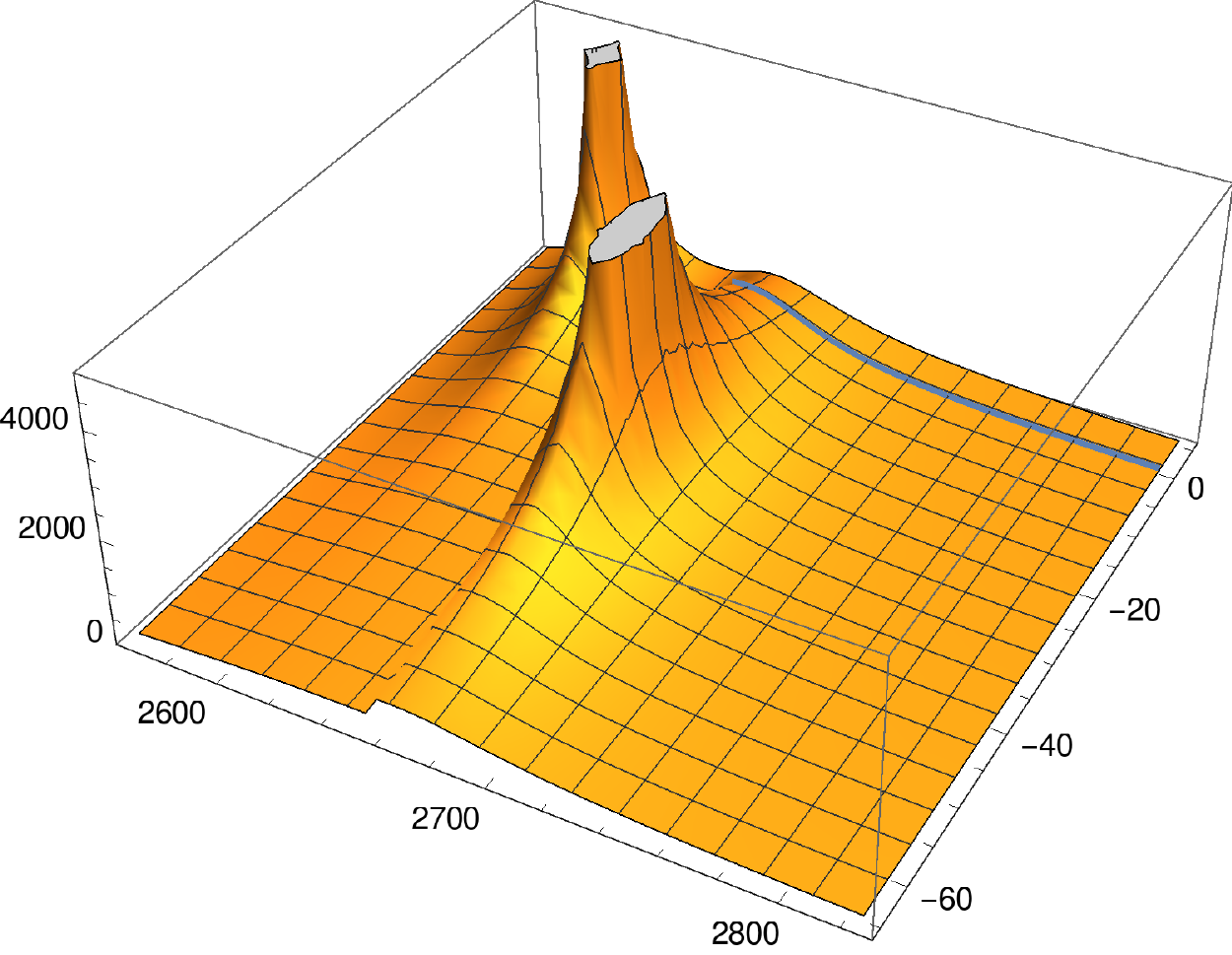}\hspace{0.75cm}\includegraphics[width=0.26\textwidth]{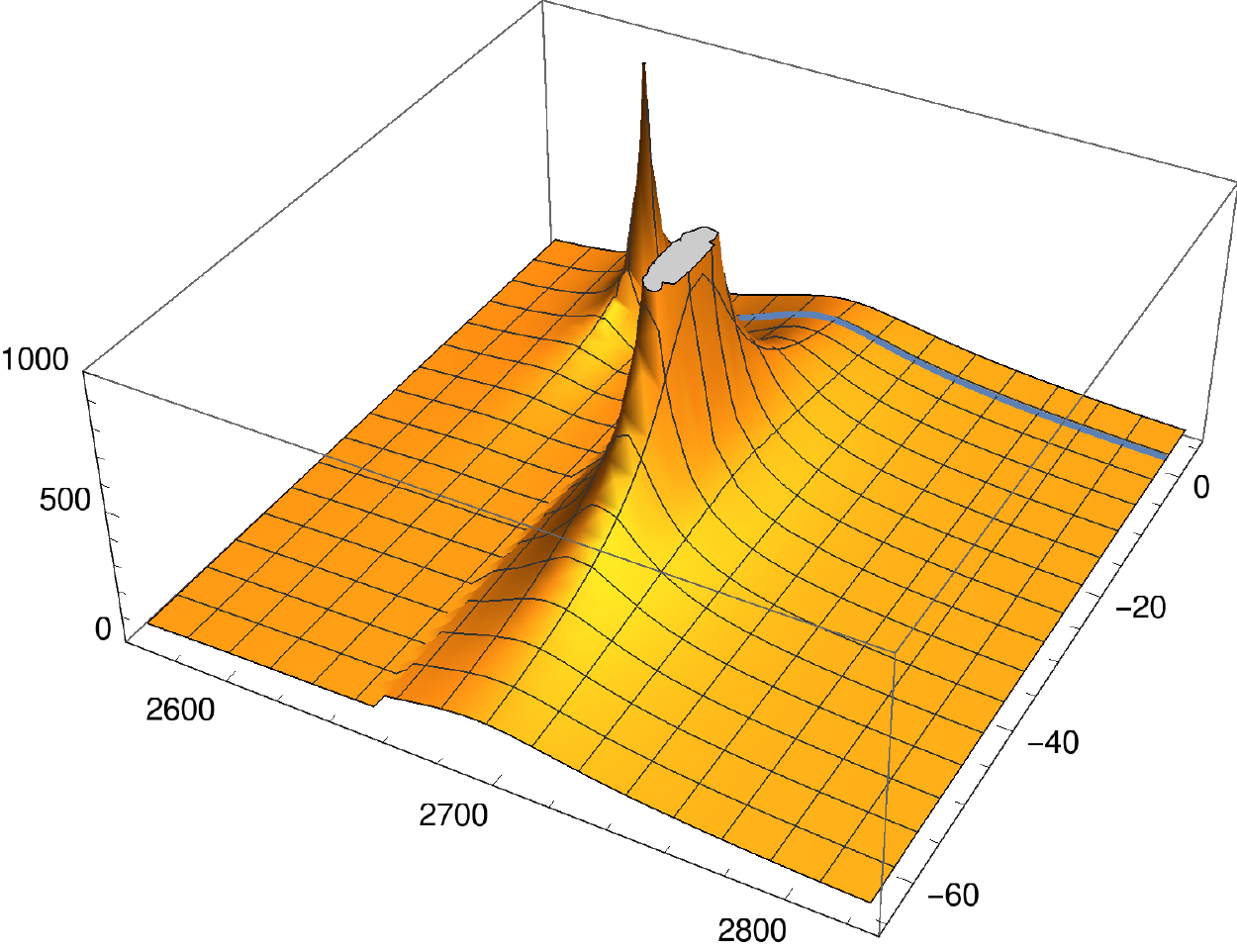}\hspace{0.75cm}\includegraphics[width=0.26\textwidth]{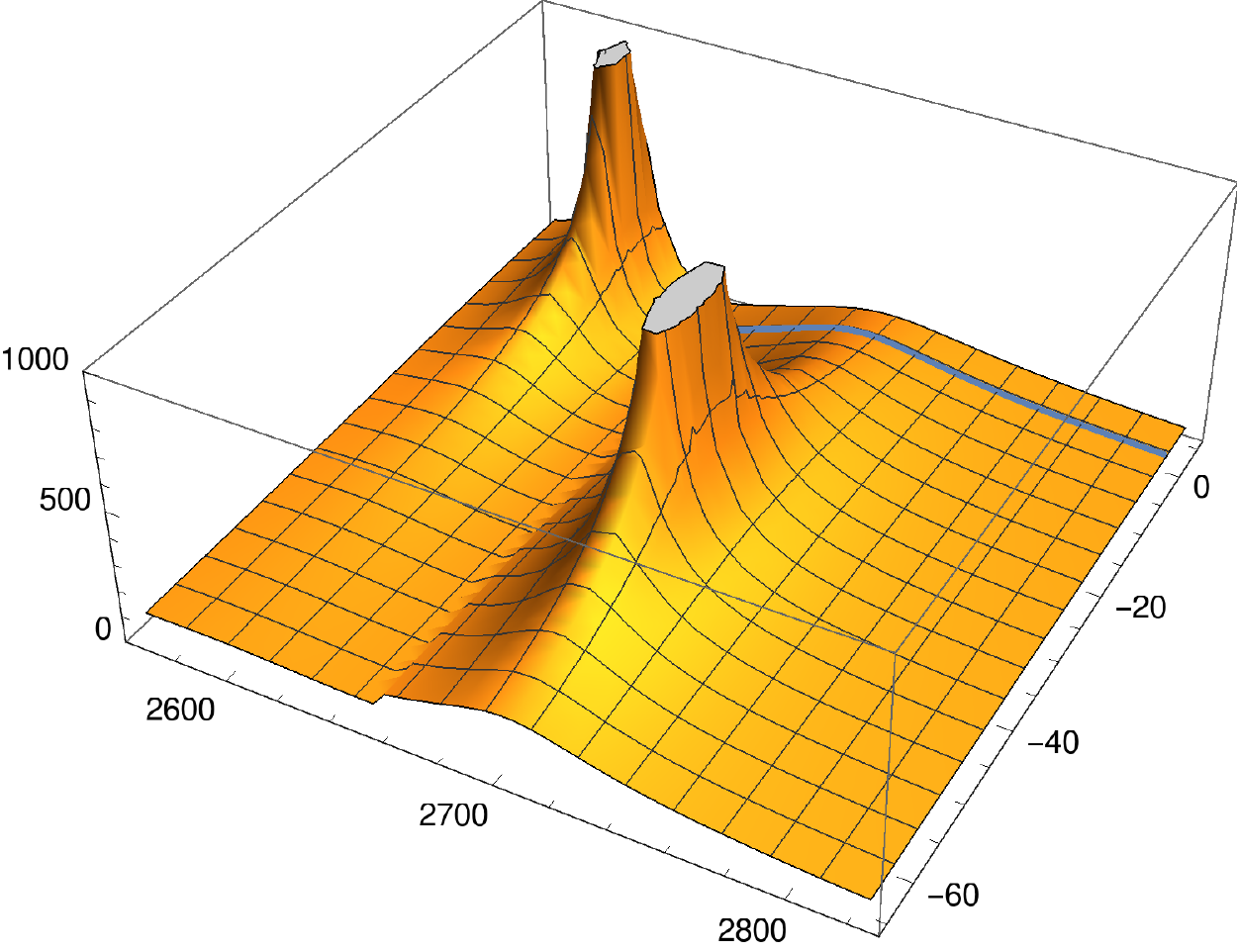}}
\end{center}
\caption{AVD-$T$ in the $J^P=3/2^-$ sector using two UV renormalization schemes :  SC$\mu$ ($\alpha=0.95$)  and a cutoff of 650 MeV in the bottom and  top panels, respectively, for different CQM \& baryon-meson pair couplings (from left to right): $(c_1=-3, d_1=-0.8)$,  $(c_1=0,  d_1=-0.8)$  and  $( c_1=2,  d_1=-0.8)$. We display the  AVD-$T$ for both the FRS (${\rm Im}(E)> 0$) and the SRS (${\rm
    Im}(E)< 0$) [fm] of the unitarized amplitudes as a function
  of the complex energy $E$ [MeV]. We also show the scattering line (blue solid curve) in all the cases. Axes are defined as in Fig.~\ref{fig:noCQM}.\label{fig:FRSSRS-32CQM}}
\end{figure}

First we pay attention to the $J^P=3/2^-$ sector. In Figs.~\ref{fig:32c1zero}, \ref{fig:32c1nozero} and \ref{fig:FRSSRS-32CQM}, we show  results obtained using SC$\mu$ ($\alpha=0.95$)  or  a common UV cutoff of 650 MeV for different CQM \& baryon-meson  couplings. In principle one expects that $d_1$ should be more relevant than $c_1$ because the $\Sigma_c^*\pi-$threshold is closer to $\Mbare$ than the $ND^*$ one. Thus, in  a first stage we set $c_1$ to zero and start varying $d_1$. Results are depicted in  Fig.~\ref{fig:32c1zero} (note that in this situation, the irreducible amplitudes depend on $d_1^2$). There are  now two poles in both renormalization schemes. The lightest one is located below the  $\Sigma_c^*\pi-$threshold and it tends to $\Mbare$ when $d_1\to 0$. Its  coupling to  $\Sigma_c^*\pi$,   $|g_{\Sigma_c^*\pi}|$,   grows  from zero, when $d_1=0$, to values of around 1.8 or 1.9, when $d_1=-2$, for the UV cutoff or SC$\mu$ renormalization schemes, respectively. The upper bound of Eq.~\eqref{eq:upper} is not satisfied above $|d_1|> 1.2 (1.0)$ for the $\Lambda=650$ MeV (SC$\mu$) scheme.

On the other hand, the second pole, located a higher masses,  is a broad resonance, with a width of around 60 MeV and little sensitivity to $d_1$. Indeed, as can be seen in the figure, the width varies less than 2 (8) MeV 
in the UV cutoff (SC$\mu$) scheme, when $d_1$ changes from 0 to $-2$. The mass of this second resonance is more affected by $d_1$, and gets bigger when $d_1^2$ increases, since the CQM exchange interaction is repulsive for energies above $\Mbare$. The pole matches the ${\rm SU(6)}_{\rm lsf}\times$HQSS molecular one discussed in Subsec.~\ref{sec:Su6results}, when the coupling between CQM and baryon-meson degrees of freedom is switched off.

Within the non-relativistic CQM used in Ref.~\cite{Arifi:2017sac}, the LEC $d_1$ is predicted to be $-0.8$. With all cautions, already mentioned, about the matching between quark-models and hadron-hadron based images of the problem, and the dependence on the renormalization procedure, we will fix $d_1$ to the latter value, and study the dependence of the previous results on $c_1$. We let this latter parameter to vary in the range $-3$ to 2 in  Fig.~\ref{fig:32c1nozero}, where the $|g_{\Sigma_c^*\pi}|$ and  $|g_{ND^* }|$ couplings, the mass and $\Sigma^*_c \pi-$molecular probability of the $J^P=3/2^-$ dressed CQM pole are shown 
as a function of $c_1$. The molecular probability is defined through the Weinberg's compositeness  rule~\cite{Weinberg:1962hj, Weinberg:1965zz}, generalized for the coupled-channels BSE formalism in Refs.~\cite{Gamermann:2009uq,Garcia-Recio:2015jsa},
\begin{equation}
 P_{\Sigma_c^*\pi}=-g^2_{\Sigma_c^*\pi}\, \left. \frac{\partial G_{\Sigma_c^*\pi}(\sqrt{s})}{\partial\sqrt{s}}\right|_{\sqrt{s}=\sqrt{s_R}}\label{eq:prob}
\end{equation}
As can be seen in the left panel of the figure, the variation of the mass of the CQM pole  with $c_1$ is quite mild within the cutoff scheme. It changes only about 3 MeV, $M_{\rm CQM}\in[2614.7, 2617.8]$ MeV, when $c_1$ varies in the  $[-2.0, 0.5]$ interval and,  at most, $M_{\rm CQM}$ reaches values close to 2630 MeV for the largest positive values of $c_1$ shown in the figure. At the same time $|g_{\Sigma_c^*\pi}|$ goes from 1.3 down to 0.8 when $c_1$ varies from $-3$ to 2. Hence, one can accommodate the experimental mass in the region of 2628 MeV, consistently with the  upper bound on the $\Sigma_c^*-$resonant contribution to the width discussed in Eq.~\eqref{eq:upper}. The molecular probability of this state would be small since $P_{\Sigma_c^*\pi}\sim 0.1$, reaching maximum values of about 0.2, when $c_1$ is close to $-3$. Moreover, for this latter value of $c_1$, $P_{ND^*}$, defined analogously to $P_{\Sigma_c^*\pi}$ in Eq.~\eqref{eq:prob} with the obvious replacements, is of the order of 0.04. When $c_1$ increases, $P_{ND^*}$ (not shown in the plot) continues decreasing and it becomes zero  close to $c_1=0.2$. From  this point, $P_{ND^*}$ starts growing to reach values of the order of 0.08 for $c_1=2$. The coupling $|g_{ND^*}|$, displayed in the figure, follows a similar pattern, as expected.

Results obtained within the SC$\mu-$scheme, shown in the right panel of Fig.~\ref{fig:32c1nozero},  differ from those discussed above, but some qualitative features are similar. The  $P_{ND^*}$ and $|g_{ND^*}|$ patterns, the small molecular probability  and the mild dependence of $M_{\rm CQM}$ and  $|g_{\Sigma_c^*\pi}|$ on $c_1$. The maximum values obtained for the mass ($\sim 2618$) of the state are found for $c_1$ in the region of $-0.8$. Note however that the possible tension with the experimental mass of 2628.11 MeV is not really significant, since the agreement can be likely improved by changing the renormalization parameter $\alpha$.

In Table~\ref{tab:j32}, we present together the properties of  CQM and molecular $J^P=3/2^-$ poles, for $d_1=-0.8$ and $c_1=0,1$ and $-1$, and both renormalization schemes. The dressed CQM  results of the table were already discussed in  Fig.~\ref{fig:32c1nozero}. The properties of the molecular state, that would have $j^\pi_{ldof}=1^-$ quantum-numbers, hardly depend on $c_1$ and both renormalization schemes predict a new state around 2.7 GeV and 60 MeV of width. The emergence of this new resonance, that would not be the $\Lambda_c(2625)$, can be clearly seen  in the FRS and SRS plots of Fig.~\ref{fig:FRSSRS-32CQM}, where larger values of $|c_1|$ than in Table~\ref{tab:j32} have been considered. 

Hence, we conclude that the physical $\Lambda_c(2625)$ finds naturally its origin in the CQM bare state obtained in Ref.~\cite{Yoshida:2015tia}, while we predict the existence of a molecular baryon, moderately broad, with a mass of about 2.7 GeV and sizable couplings to both $\Sigma_c^* \pi$ and $ND^*$.

This latter pole will not show up in the experimental $\Lambda_c\pi\pi$ spectrum, dominated by the physical resonance. Furthermore this state, mistakenly associated with the $\Lambda_c(2625)$ in the previous ${\rm SU(6)}_{\rm lsf}\times$HQSS studies of 
Refs.~\cite{GarciaRecio:2008dp, Romanets:2012hm} where the coupling to CQM degrees of freedom was not considered,  will be similar to that found in the chiral approach of Ref.~\cite{Hofmann:2006qx} or to the $\Sigma_c^*\pi$ pole reported in the ELHG scheme followed\footnote{In that work,   
it was not identified with the $\Lambda_c(2625)$ resonance, which is generated there as a $ND^*$ state, after modifying the ELHG $N D^*\to  N D^*$ potential including 
box-diagrams constructed out of the anomalous $D^*D^*\pi$ coupling, and fitting the UV cutoffs to reproduce its mass.} in \cite{Liang:2014kra}.  The SU(3) chiral approach of 
Ref.~\cite{Lu:2014ina} reduces the mass of this molecular state down to that of the $\Lambda_c(2625)$ by using a large UV cutoff of 2.13 GeV. This points out, following the arguments given in ~\cite{Guo:2016nhb, Albaladejo:2016eps}, to the existence of some relevant degrees of freedom  (CQM states and/or $ND^{*}$ components) that are not properly accounted for in \cite{Lu:2014ina}.

\subsubsection{The $\Lambda_c(2595)$}
\begin{figure}[h]
\begin{center}
\makebox[0pt]{\includegraphics[width=0.5\textwidth]{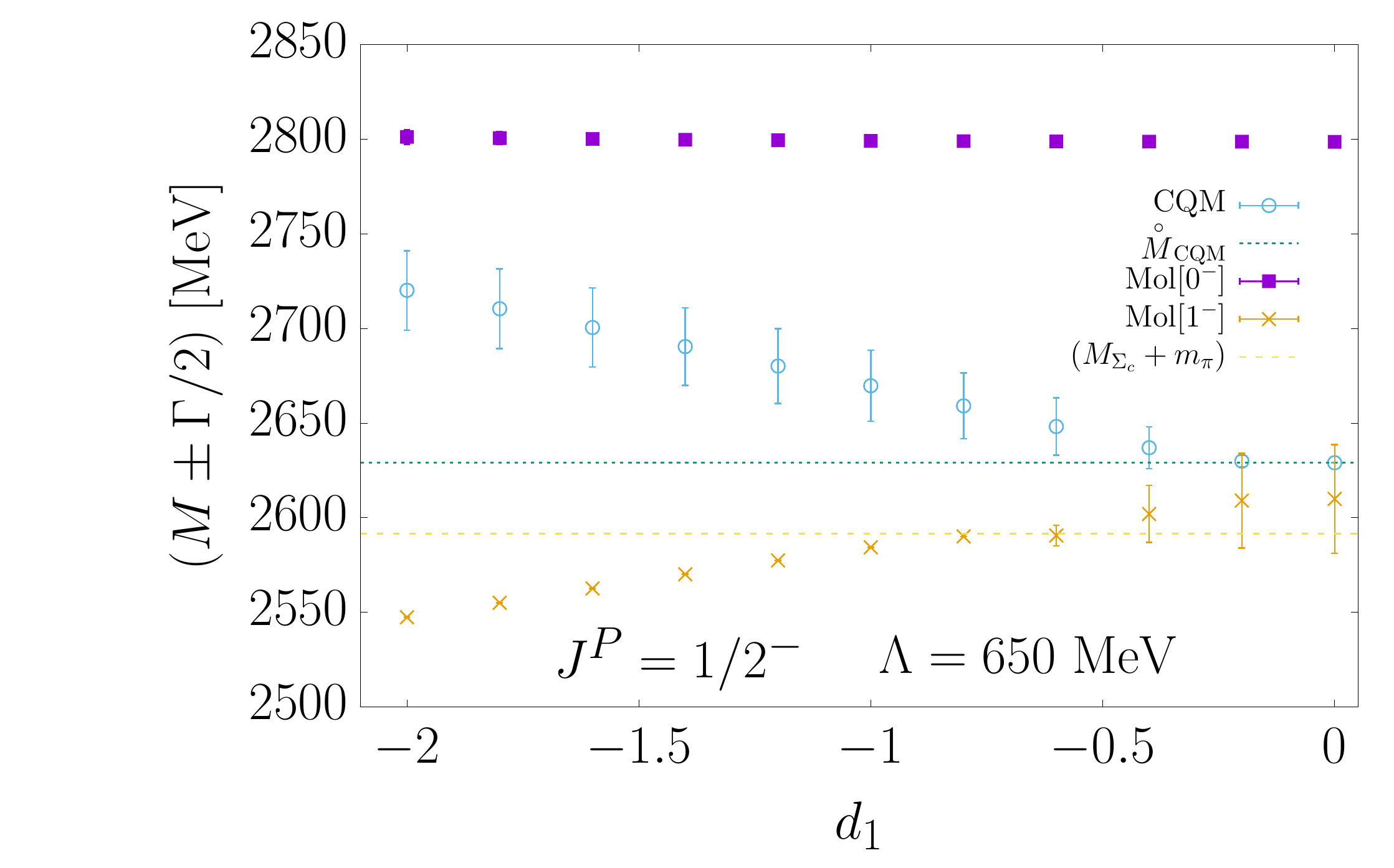}\hspace{0.25cm}\includegraphics[width=0.5\textwidth]{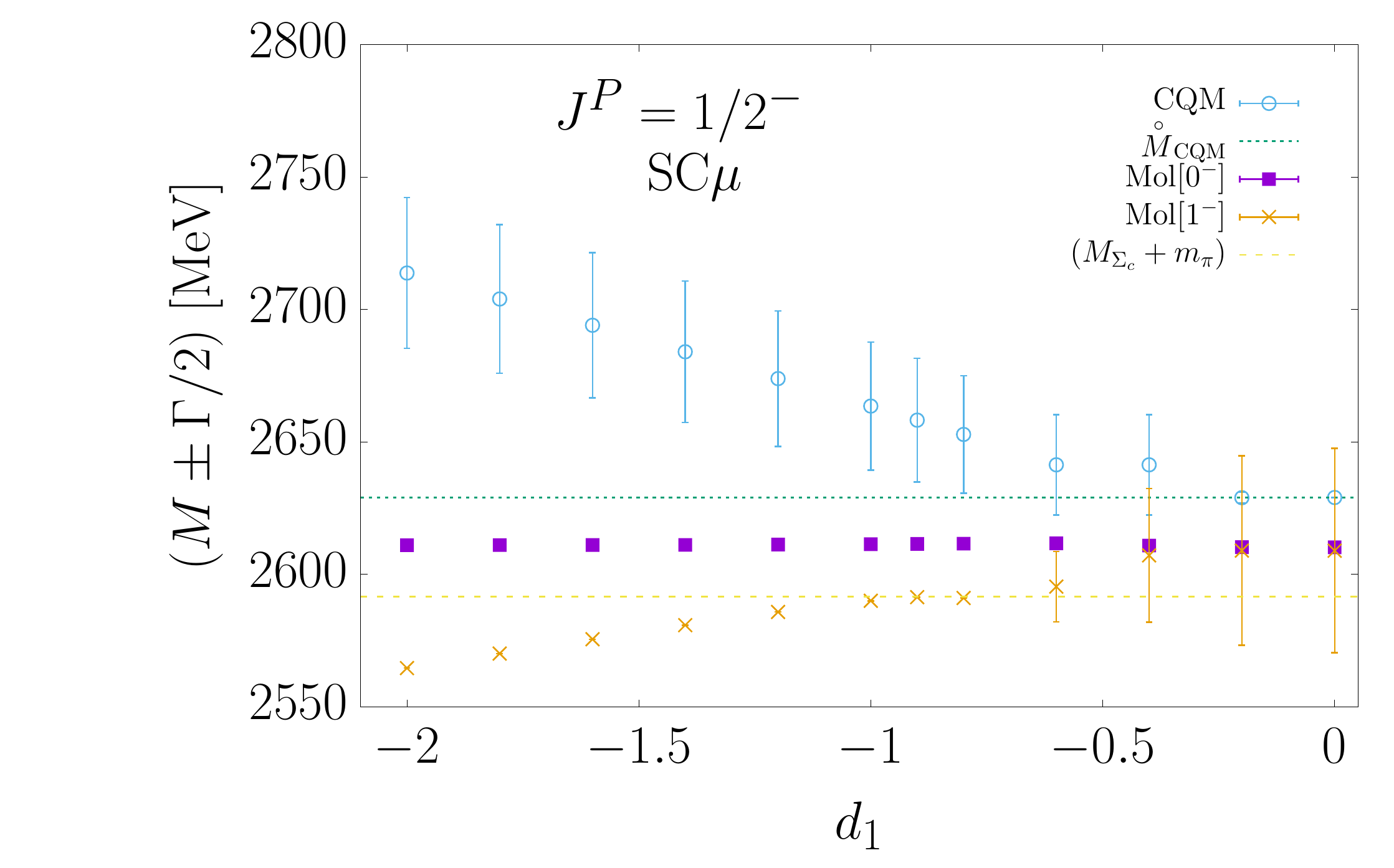}}
\end{center}
\caption{Dependence of the $J^P=1/2^-$ CQM and molecular pole positions as a function of the LEC $d_1$, for $c_1=0$. We show results for both, the cutoff and SC$\mu$ ($\alpha=0.95$) renormalization schemes, and the values of the bare CQM mass and the $\Sigma_c\pi$ threshold energy. Molecular states are labeled according to their dominant {\it ldof} configuration, $0^-$ or $1^-$.   }\label{fig:12c1zero}
\end{figure}
\begin{figure}[h]
\begin{center}
\makebox[0pt]{\includegraphics[width=0.26\textwidth]{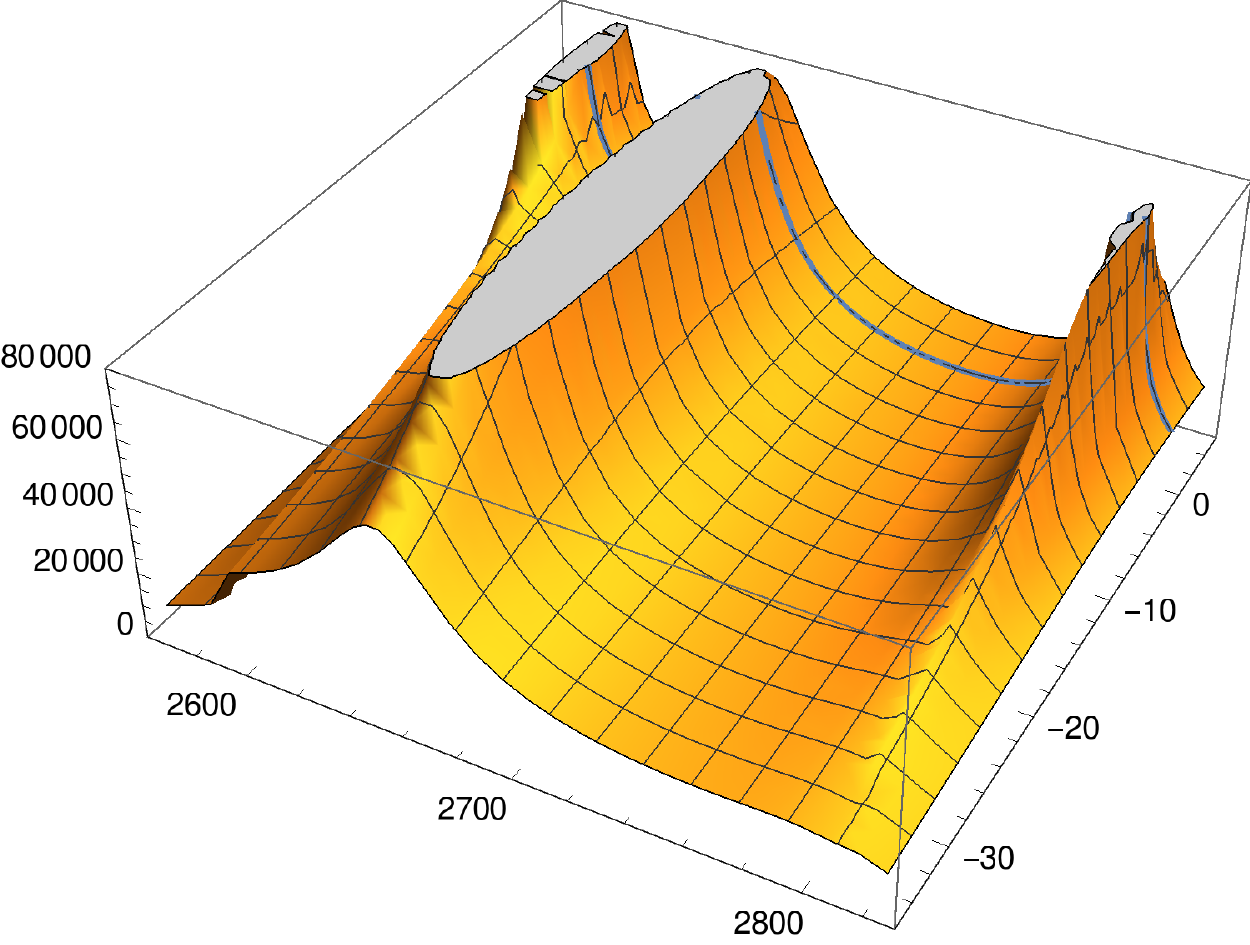}\hspace{0.75cm}\includegraphics[width=0.26\textwidth]{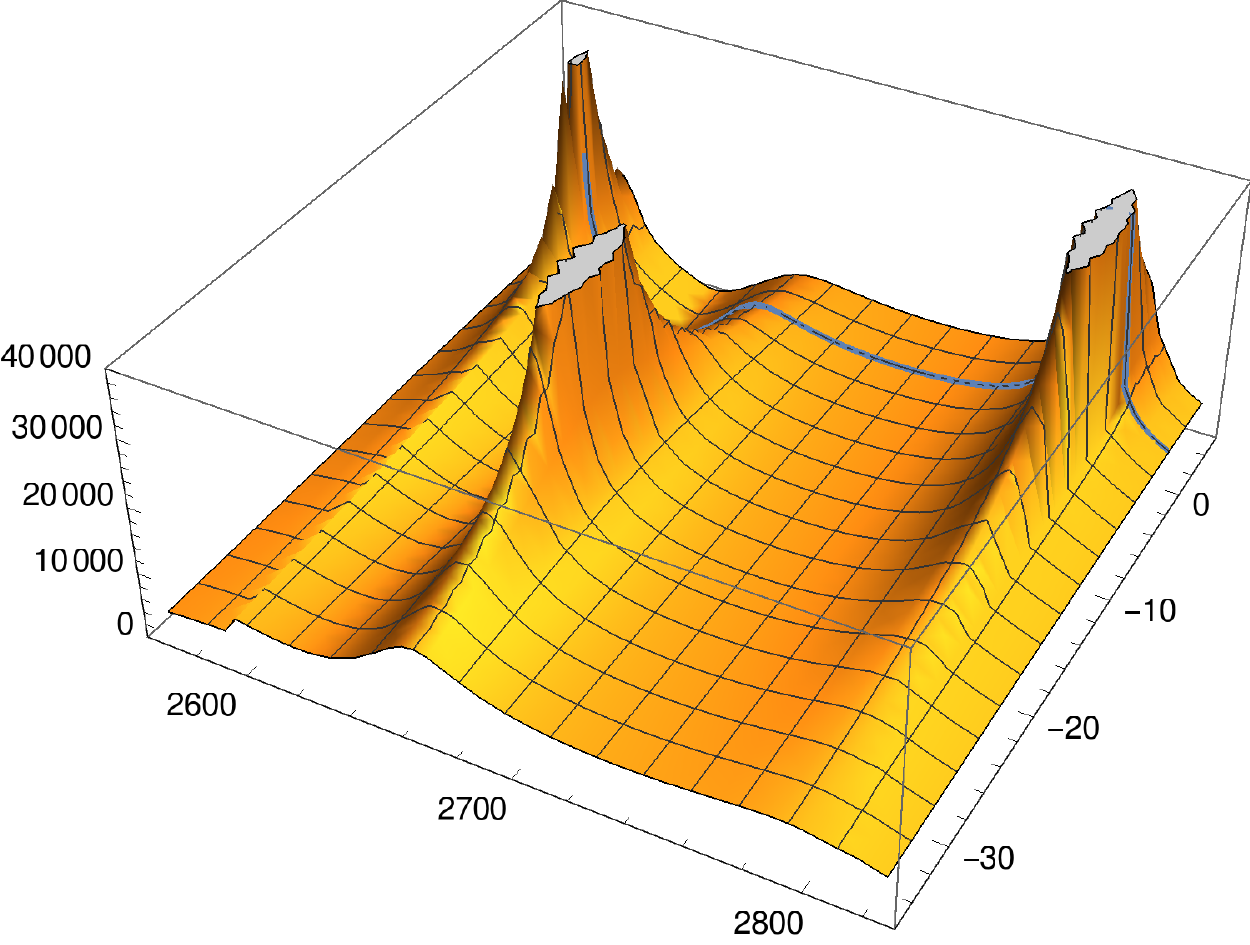}\hspace{0.75cm}\includegraphics[width=0.26\textwidth]{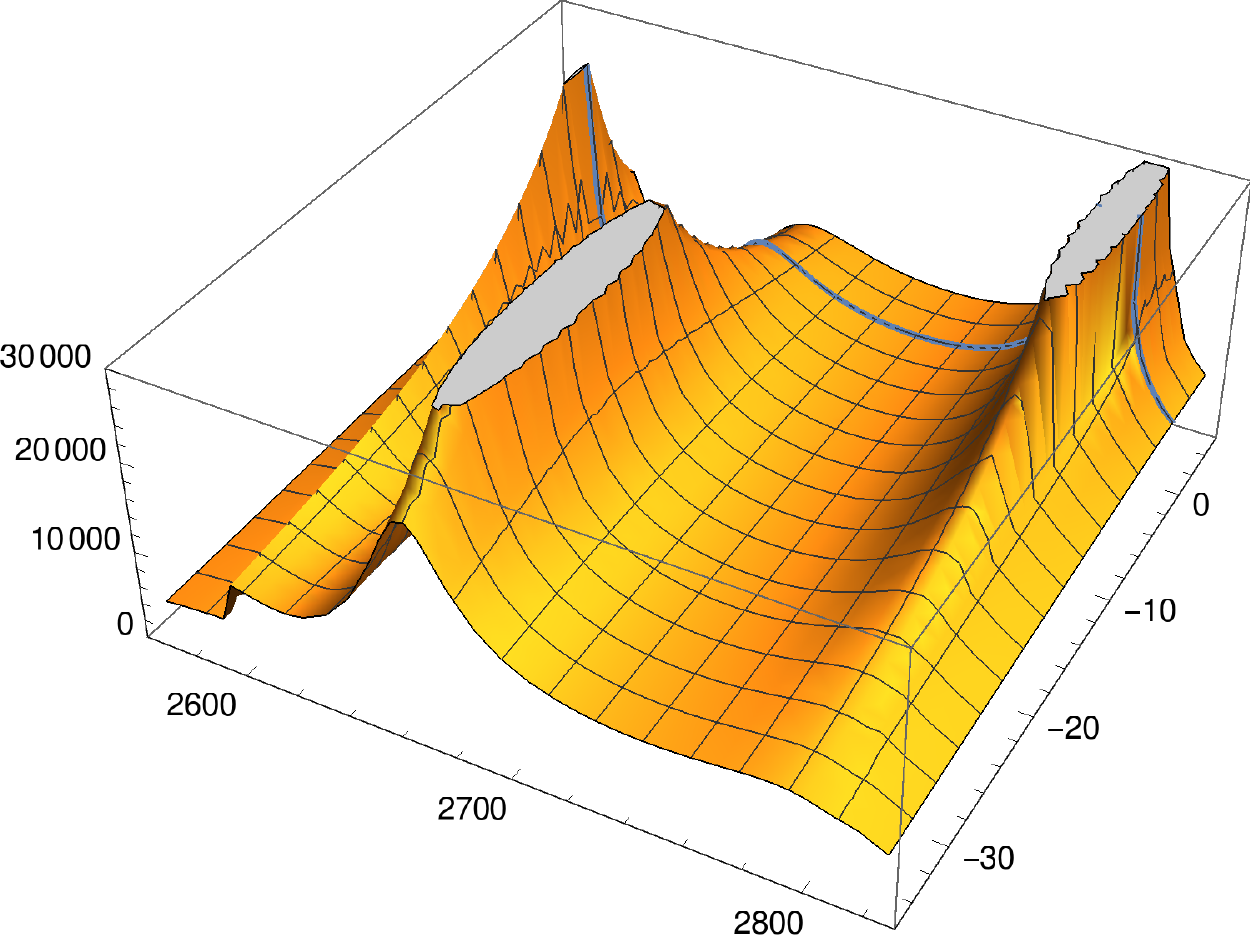}}\\\vspace{0.5cm}
\makebox[0pt]{\includegraphics[width=0.26\textwidth]{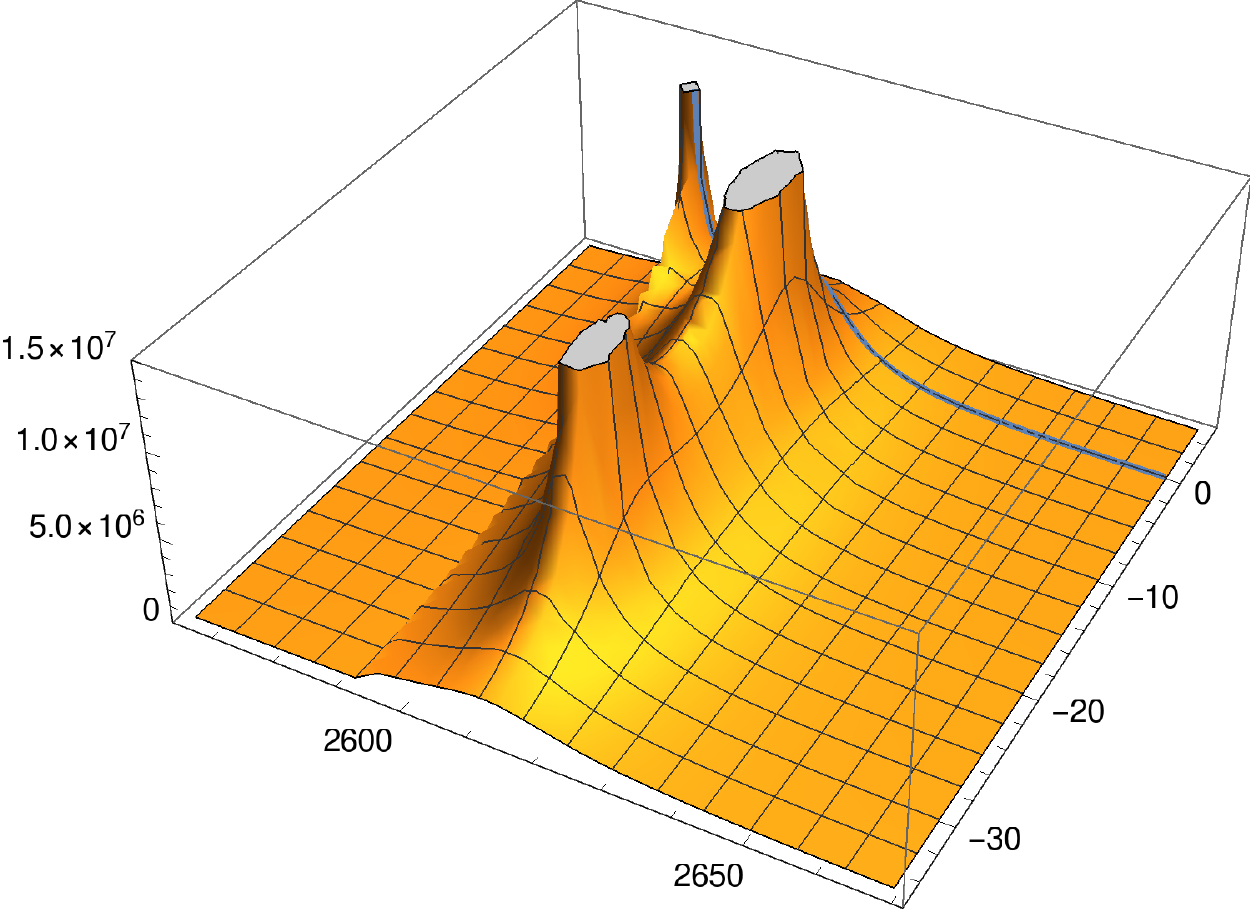}\hspace{0.75cm}\includegraphics[width=0.26\textwidth]{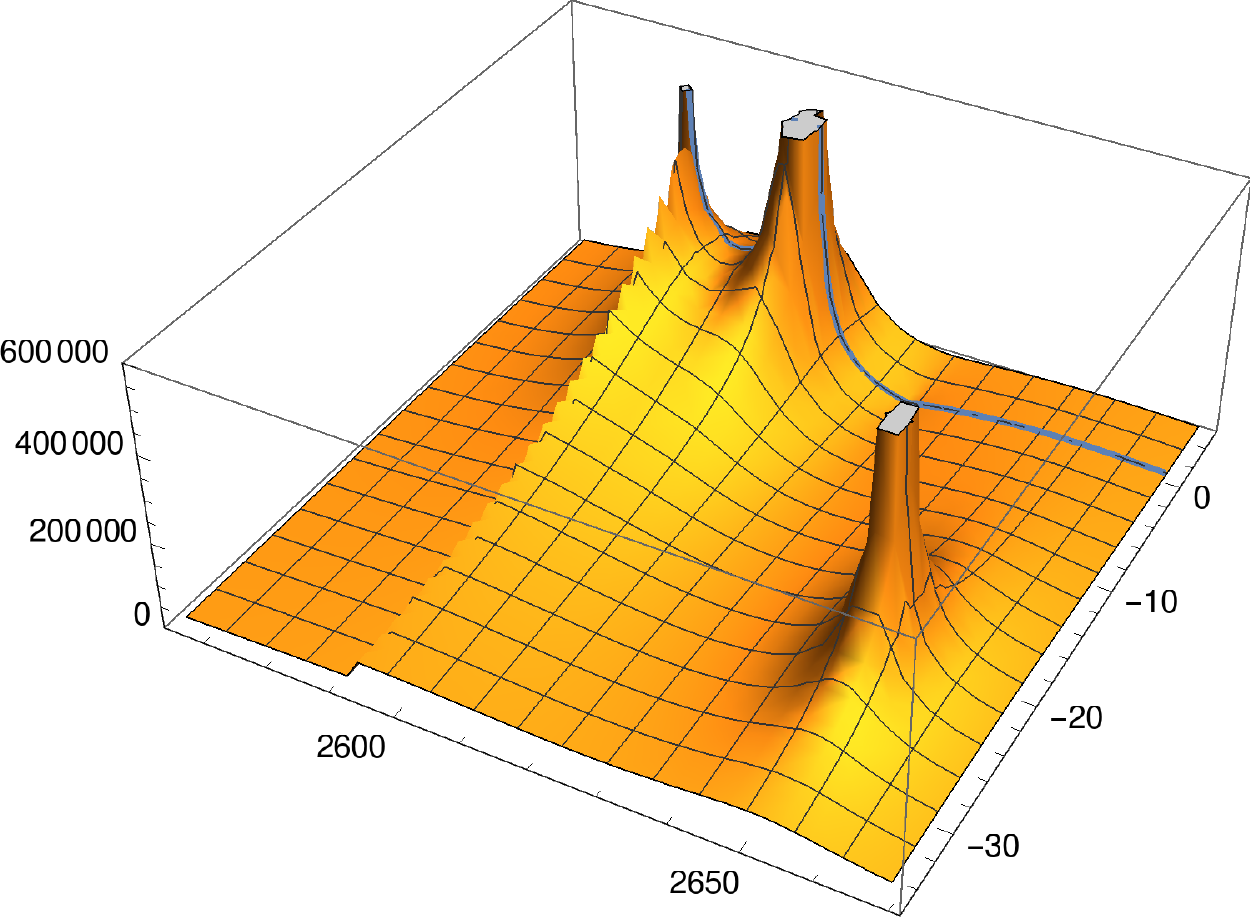}\hspace{0.75cm}\includegraphics[width=0.26\textwidth]{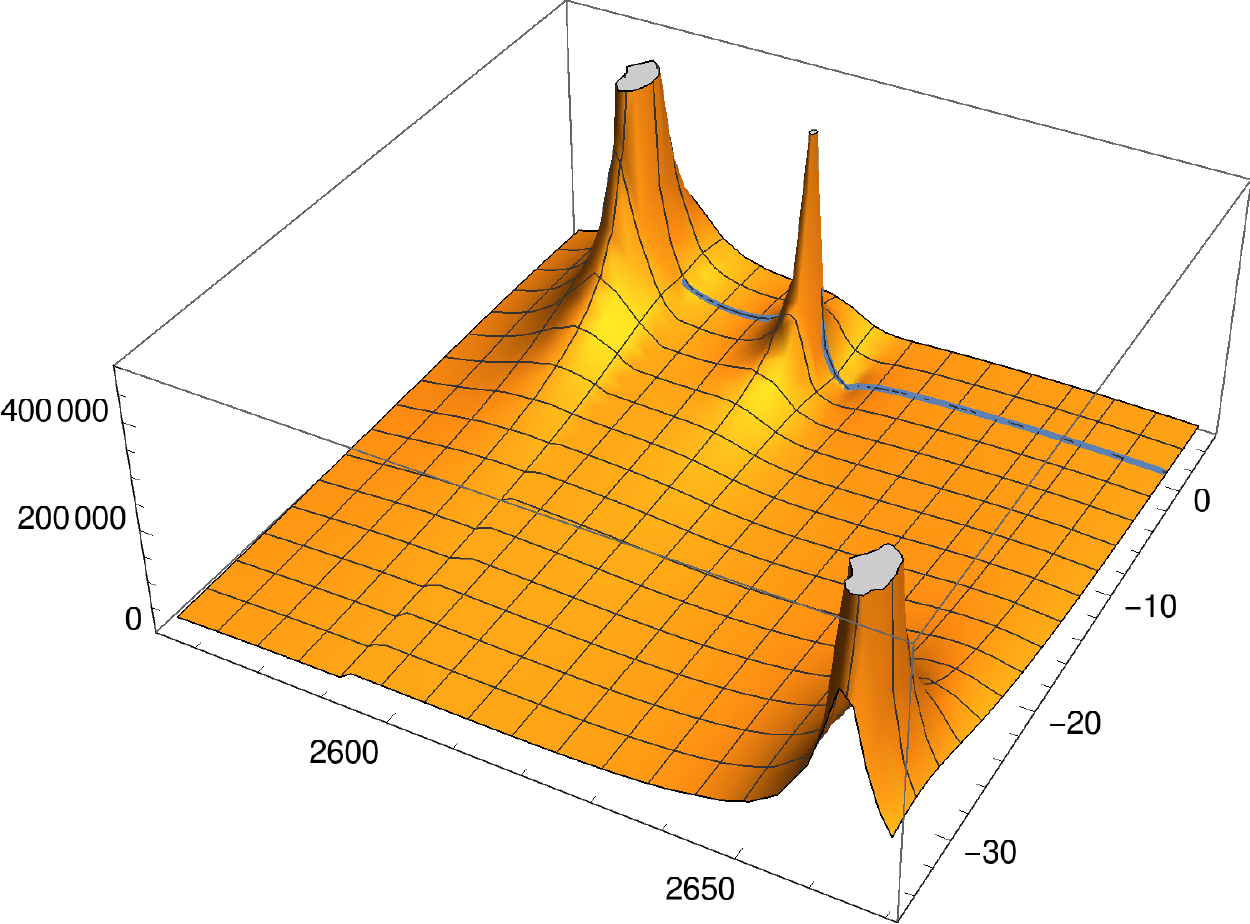}}
\end{center}
\caption{AVD-$T$ in the $J^P=1/2^-$ sector using two UV renormalization schemes :  SC$\mu$ ($\alpha=0.95$)  and a cutoff of 650 MeV in the bottom and  top panels, respectively, for different CQM \& baryon-meson pair couplings (from left to right): $( c_1=-3,  d_1=-0.8)$,  $( c_1=0,  d_1=-0.8)$  and  $( c_1=2,  d_1=-0.8)$. We display the  AVD-$T$ for both the FRS (${\rm Im}(E)> 0$) and the SRS (${\rm
    Im}(E)< 0$) [fm] of the unitarized amplitudes as a function
  of the complex energy $E$ [MeV]. We also show the scattering line (blue solid curve) in all the cases. Note that the range of  ${\rm Re}(E)$ is much larger in the top panels than in the bottom ones. Axes are defined as in Fig.~\ref{fig:noCQM}.}~\label{fig:FRSSRS-12CQM}
\end{figure}
Now, we turn the discussion into the $J^P = 1/2^-$ sector. As we did before, in a first stage we set $c_1$ to zero and start varying $d_1$. Results are depicted in  Fig.~\ref{fig:12c1zero}. There appear now three poles for both renormalization schemes considered in this work. As compared to the case in the left panels of Fig.~\ref{fig:noCQM}, where  the coupling between CQM and baryon-meson degrees of freedom was switched off, there is an extra state which has its origin in the $j^\pi_{ldof}=1^-$ CQM bare state. The mass and the width of the narrow state at 2800 MeV ($\Lambda=650$ MeV) or 2610 MeV (SC$\mu$) are practically unaltered by $d_1$. This is a trivial consequence of the largely dominant $j_{ldof}^\pi=0^-$ configuration of these states, since HQSS forbids their coupling  to the $j^\pi_{ldof}=1^-$ CQM bare state. 

The location of the second broad molecular state, [$\Lambda_{c\, (b)}^{\rm 1^-}(2595)$], observed in Fig.~\ref{fig:noCQM} is strongly influenced by the quark-model state that produces an attraction that grows with $d_1^2$. Thus, for $d_1< -0.6$ or $-0.7$, depending of the renormalization procedure, 
it moves below the $\Sigma_c\pi$ threshold and becomes a bound state. Within the SC$\mu$ scheme, this $j^\pi_{ldof}=1^-$ molecular state would not be, however, identified with the physical $\Lambda_c(2595)$ resonance, that would be reproduced by the narrow $\Lambda_{c\, (n)}^{\rm 0^-}(2595)$ pole at 2610-2611 MeV, with small $|g_{\Sigma_c\pi}|$ coupling\footnote{It decreases with $d_1^2$ and it varies from 0.5 for $d_1=0$ down to 0.04 for $d_1=-2$. } and large $|g_{ND}|$ and $|g_{ND^*}|$ ones, especially the latter ($\ge 6.2$). The situation is different in the UV cutoff scheme, since the $j^\pi_{ldof}=0^-$ narrow resonance  is placed at 2800 MeV, and it is precisely the  $j^\pi_{ldof}=1^-$ molecular state, the best candidate to describe the physical $\Lambda_c(2595)$. 

In addition, we see in Fig.~\ref{fig:12c1zero} that the bare CQM state is modified due to the baryon-meson loop
effects, and it is moved to the complex plane acquiring also a finite width that obviously grows with $d_1^2$.  The quantitative details, nevertheless,  depend on renormalization scheme.

As in the $\Lambda_c(2625)$ subsection, we  fix $d_1=-0.8$ from the CQM of Ref.~\cite{Arifi:2017sac}, and study in  
Table~\ref{tab:j12}  and Fig.~\ref{fig:FRSSRS-12CQM}, the  dependence of the spectrum of states on the LEC $c_1$. As expected, the mass position of the $j^\pi_{ldof}=0^-$ pole is hardly affected, while its small width depends much more on $c_1$. As mentioned above, within the SC$\mu$ scheme, the physical $\Lambda_c(2595)$ is identified with the $\Lambda_{c\, (n)}^{\rm 0^-}(2595)$. We see that the pole might have a coupling to the $\Sigma_c\pi-$pair smaller than 1,  and thus it would be  smaller than needed to reproduce the experimental width from Eq.~\eqref{eq:resulwdth}.  In the UV cutoff approach, instead, there would be a molecular narrow state close to the $ND$ threshold, strongly coupled to it and that might provide some visible signatures in processes involving final state interactions of this baryon-meson pair (see bottom panels of Fig.~\ref{fig:FRSSRS-12CQM}). In this latter renormalization scheme, the $\Lambda_c(2595)$ is described by the $j^\pi_{ldof}=1^-$ hadron molecule located below threshold at around 2590 MeV, little affected by $c_1$,  and with  $|g_{\Sigma_c\pi}|~\sim 1$. Thus from   Eq.~\eqref{eq:resulwdth},  the $\Lambda_c(2595)\to \Lambda \pi \pi$ width will be predicted to be around 1.8, in good agreement with experiment ($2.6 \pm 0.6$). Nevertheless, the $\Lambda_c(2595)$, despite of having $1^-$ quantum numbers for the {\it ldof}, would not be the HQSS partner of the  $\Lambda_c(2625)$  either in this case, because the predominantly quark-model structure of the latter. Indeed, the  $\Lambda_c(2595)$ would have a large molecular content, $P_{\Sigma_c^*\pi}=0.6-0.7$.  

Note that the $j_{\it ldof}^\pi = 1^-$state in the SC$\mu$ scheme will be irrelevant, since its effects will be completely overcome by those produced by the $\Lambda_{c\, (n)}^{\rm 0^-}(2595)$ (see Fig.~\ref{fig:FRSSRS-12CQM}), independently it is placed below the $\Sigma_c\pi$ threshold or it becomes a broad resonance. 

The different inner structure of the $\Lambda_c(2595)$ within the UV cutoff and SC$\mu$ schemes, and the dependence of this structure on $c_1$ will lead to differences in $ND$ and $ND^*$ couplings that would produce different predictions for the exclusive semileptonic $\Lambda_b\to \Lambda_c(2595)$ decay~\cite{Liang:2016exm,Nieves:2019kdh}. 

Finally, we see that in both renormalization schemes we obtain the dressed CQM pole at masses around 2640--2660 MeV and with a width of the order of 30-50 MeV, depending on the chosen regulator and on  $c_1$, though for moderate variations, one should not expect a large dependence on $c_1$ because the $ND$ and $ND^*$ thresholds are not too close. This is a prediction of the present work, and this state should provide signatures in the open channel $\Sigma_c\pi$ since its coupling to this pair is sizable, well above one. As seen in Fig.~\ref{fig:FRSSRS-12CQM}, for large negative values\footnote{What it is really relevant is that the product $c_1\,d_1$ is positive.} of $c_1$ in the SC$\mu$ case, it could be, however, be shadowed by the  $\Lambda_{c\, (n)}^{\rm 0^-}(2595)$ resonance. 

\section{Conclusions}
\label{sec:concl}

We have shown that the $\Lambda_c(2595)$ and the $\Lambda_c(2625)$ are not HQSS partners. The $J^P=3/2^-$ resonance should be viewed mostly as a quark-model state naturally predicted to lie  very close to its nominal mass~\cite{Yoshida:2015tia}. This contradicts a large number of molecular scenarious suggested for this resonance in the literature. In addition, there will exist a molecular baryon, moderately broad, with a mass of about 2.7 GeV and sizable couplings to both $\Sigma_c^* \pi$ and $ND^*$, that will fit into the expectations of being a $\Sigma^*_c\pi$ molecule generated by the chiral interaction of this pair. 

The $\Lambda_c(2595)$ is predicted, however, to have a predominant molecular structure. This is because, it  is either the result of the chiral $\Sigma_c\pi$ interaction, which threshold is located much more closer than the mass of the bare three-quark state, or because the {\it ldof} in its inner structure are coupled to the unnatural $0^-$ quantum-numbers. The latter is what happens in the SC$\mu$ renormalization scheme that enhances the influence of the $N D^*$ channel in the dynamics of this narrow resonance. Attending to the three-body $\Lambda_c(2595)\to \Lambda \pi \pi$ decay width, the SC$\mu$ scenarious is slightly disfavored, and it looks more natural to assign a $1^-$ configuration to the {\it ldof} content of 
the physical $\Lambda_c(2595)$ state, as found when an UV cutoff is employed. 

We also obtain a further $J^P=1/2^-$ resonance that is the result of dressing with  baryon-meson loops the bare CQM pole. 
It would have a mass of around 2640--2660 MeV and  a width of the order of 30-50 MeV. Finally, within the UV cutoff renormalization scheme, we also find a narrow state at 2800 MeV close to the $ND$ threshold. This state  has large $ND$ and $ND^*$ couplings and it should provide some visible signatures in processes involving final state interactions of the $ND$ and $ND^*$ pairs.

The spectrum  found in this work cannot be easily understood in terms of HQSS, despite having  used interactions that respect this symmetry. This is because  the bare quark-model state and the $\Sigma_c\pi$ threshold are located extraordinarily close to the $\Lambda_c(2625)$ and $\Lambda_c(2595)$, respectively, and hence they play totally different roles in each sector. Note that $(M_{\Sigma_c^*}-M_{\Sigma_c})\sim 65$ MeV is around a factor of two larger than the $\Lambda_c(2625)-\Lambda_c(2595)$ mass splitting. This does not fit well into a molecular picture of these two resonances generated by $\Sigma_c^{(*)}\pi$ chiral forces, and in addition  the splitting found in the CQM study of Ref.~\cite{Yoshida:2015tia} is only of 2 MeV, much smaller than any of the mass differences quoted above.     Moreover, the SC$\mu$ renormalization scheme leads to an unexpected enhancing of the importance of the $j_{ldof}^\pi=0^-$ components of the ${\rm SU(6)}_{\rm lsf}\times$HQSS interaction in the $J^P=1/2^-$ sector, which are  driven by $ND-ND^*$ coupled-channels interactions. This is not the case when an UV cutoff is employed.   

\section*{Acknowledgments}
R.P.~Pavao wishes to thank the program Santiago Grisolia of the
Generalitat Valenciana. This research has been supported by the
Spanish Ministerio de Ciencia, Innovaci\'on y Universidades and
European FEDER funds under Contracts FIS2017-84038-C2-1-P and
SEV-2014-0398 and by the EU STRONG-2020 project under the program
H2020-INFRAIA-2018-1, grant agreement no. 824093.

\bibliography{biblio}
\end{document}